\documentclass[pre,aps,twocolumn,superscriptaddress]{revtex4}

\usepackage[dvips]{graphicx}
\usepackage{amssymb,amsfonts,amsmath}
\usepackage{color}
\usepackage{ulem}
\usepackage[hidelinks]{hyperref}
\usepackage{bbm}
\usepackage{bm}
\usepackage{xcolor}

\newcommand{\rv}{{\mathbf r}}

\newcommand{\Tr}{{\rm Tr}\,}

\newcommand{\pv}{{\bf p}}

\newcommand{\Fv}{{\bf F}}

\newcommand{\eps}{{\boldsymbol \epsilon}}
\newcommand{\unity}{{\mathbbm 1}}
\newcommand{\cov}{{\rm cov}}

\newcommand{\avg}[1]{\Big\langle #1 \Big\rangle}
\newcommand{\eqr}[1]{Eq.~\eqref{#1}}
\newcommand{\ff}{{f\!f}}
\newcommand{\gradf}{{\nabla\!f}}

\newcommand{\mydelete}[1]{{}}
\newcommand{\taub}{{\boldsymbol\tau}}
\newcommand{\aas}{{\alpha\alpha'}}
\newcommand{\as}{{\alpha'}}
\newcommand{\rmint}{{\rm int}}
\newcommand{\rmext}{{\rm ext}}
\newcommand{\rmself}{{\rm self}}
\newcommand{\alphas}{{\alpha'}}
\newcommand{\hatFva}{\hat\Fv_\alpha}
\newcommand{\hatFvas}{\hat\Fv_{\alpha'}}
\newcommand{\Fva}{\Fv_\alpha}
\newcommand{\bsig}{\boldsymbol\sigma}
\newcommand{\calN}{{\cal N}}
\newcommand{\Sv}{{\bf S}}
\newcommand{\glob}{\circ}

\begin{document}

\title{Gauge invariance and hyperforce correlation theory for
  equilibrium fluid mixtures}

\author{Joshua Matthes\footnote{Authors contributed equally.}}
\affiliation{Theoretische Physik II, Physikalisches Institut, 
  Universit{\"a}t Bayreuth, D-95447 Bayreuth, Germany}

\author{Silas Robitschko$^*$}
\affiliation{Theoretische Physik II, Physikalisches Institut, 
  Universit{\"a}t Bayreuth, D-95447 Bayreuth, Germany}

\author{Johanna M\"uller}
\affiliation{Theoretische Physik II, Physikalisches Institut, 
  Universit{\"a}t Bayreuth, D-95447 Bayreuth, Germany}

\author{Sophie Hermann}
\address{CNRS, Sorbonne Universit\'e, Physicochimie des Electrolytes et 
  Nanosyst\`emes Interfaciaux, F-75005 Paris, France.}

\author{Florian Samm\"uller}
\affiliation{Theoretische Physik II, Physikalisches Institut, 
  Universit{\"a}t Bayreuth, D-95447 Bayreuth, Germany}

\author{Matthias Schmidt}
\affiliation{Theoretische Physik II, Physikalisches Institut, 
  Universit{\"a}t Bayreuth, D-95447 Bayreuth, Germany}
\email{Matthias.Schmidt@uni-bayreuth.de}

\date{23 September 2025, revised version: 27 November 2025}

\begin{abstract}
  We formulate gauge invariance for the equilibrium statistical
  mechanics of classical multi-component systems.  Species-resolved
  phase space shifting constitutes a gauge transformation which we
  analyze using Noether's theorem and shifting differential operators
  that encapsulate the gauge invariance. The approach yields exact
  equilibrium sum rules for general mixtures. Species-resolved gauge
  correlation functions for the force-force and force-gradient pair
  correlation structure emerge on the two-body level. Exact 3g-sum
  rules relate these correlation functions to the spatial Hessian of
  the partial pair distribution functions.  General observables are
  associated with hyperforce densities that measure the covariance of
  the given observable with the interparticle, external, and diffusive
  partial force density observables. Exact hyperforce and Lie algebra
  sum rules interrelate these correlation functions with each
  other. The practical accessibility of the framework is demonstrated
  for binary Lennard-Jones mixtures using both adaptive Brownian
  dynamics and grand canonical Monte Carlo simulations. Specifically,
  we investigate the force-force pair correlation structure of the
  Kob-Andersen bulk liquid and we show results for representative
  hyperforce correlation functions in Wilding {\it et~al.}'s
  symmetrical mixture confined between two asymmetric planar parallel
  walls.
\end{abstract}

\maketitle

\section{Introduction}
\label{SECintroduction}
Soft matter consists naturally of several different microscopic
components \cite{evans2019physicsToday, hansen2013}, with ions in
electrolytes \cite{cats2021decay, minh2023faraday, minh2023noise,
  cox2020pnas, bui2024, bui2024neuralrpm, bui2025electromechanics,
  bui2025dielectrocapillarity} and differently-sized colloids in glass
forming mixtures being prominent examples for the diverse range of
systems that display a wide variety of physical effects.  Targeting
specific phenomena often requires bespoke treatment. In particular the
glass formation phenomenon has been studied on the basis of a plethora
of order parameters, including measures of non-ergodicity \cite{kob1995} and point-to-set length scales \cite{hocky2013}, as well as via machine learning
\cite{cubuk2015mlGlass} and analyzing structural
motifs~\cite{malins2013}. A common observation in this realm is the
similarity of the liquid and glass states when analyzed on the pair
correlation level, as expressed succinctly by the authors of
Ref.~\cite{turci2017prx} who note that ``structural changes appear to
be minor when looking at two-point measures like the structure factor,
[while] higher-order measures reveal a richer behavior''.  A
comparison of results for pair distribution functions of different
microscopic glass forming models is presented in
Ref.~\cite{hocky2013}. Going beyond the pair distribution function,
and its species-labelled generalization to {\it partial} pair
distribution functions that characterize mixtures, is often useful.

Noether's theorem \cite{noether1918, byers1998} was applied in a
variety of different settings in statistical physics
\cite{baez2013markov, marvian2014quantum, sasa2016, sasa2019,
  revzen1970, baez2020bottom, bravetti2023}. The theorem provides the
basis for the recent thermal invariance theory
\cite{hermann2021noether, hermann2022topicalReview,
  hermann2022variance, hermann2022quantum, tschopp2022forceDFT,
  sammueller2023whatIsLiquid, hermann2023whatIsLiquid,
  robitschko2024any, mueller2024gauge, mueller2024whygauge,
  mueller2024dynamic, rotenberg2024spotted,
  miller2025physicsToday}. This approach is based on a rigorous
invariance of equilibrium averages and of thermodynamic potentials
against specific shifting and rotation operations, as described below
in detail. Force and torque correlation functions emerge
systematically within the framework and these are interrelated by
exact statistical mechanical identities (``sum rules''). The sum rules
take on the role that conservation laws play in conventional uses of
the Noether theorem, where typically the invariances within a
dynamical description are analysed.

The statistical mechanical gauge invariance gives rise to force-force
and force-gradient two-body correlation functions that reveal much
insight into the bulk structure of liquids and more general soft
matter systems \cite{sammueller2023whatIsLiquid,
  hermann2023whatIsLiquid}. Thereby the spatially resolved force-force
correlation function is crucial and distinct from the temporal force
autocorrelation function of tagged particle motion~\cite{zwanzig2001},
see e.g.~Ref.~\cite{krueger2010shear} for a study of the effects of
shear.  Here the force-force correlation function rather measures the
covariance of the forces that act on each particle in an interacting
pair. Similarly, the force-gradient correlation function represents
the mean gradient of the force that acts on one of the particles upon
displacing the second particle.  The quantitative analysis of these
gauge correlation functions allows one to trace a broad range of
microscopic structuring effects, from clear signatures of
interparticle attraction to chain formation in gels and orientational
order in liquid crystals \cite{sammueller2023whatIsLiquid,
  hermann2023whatIsLiquid}.

The theoretical structure emerges from an inherent gauge invariance of
statistical mechanics against phase space shifting
\cite{mueller2024gauge, mueller2024whygauge,
  mueller2024dynamic}. Popular accounts have been
given~\cite{rotenberg2024spotted, miller2025physicsToday} and a
dynamical generalization was presented very
recently~\cite{mueller2024dynamic}.  Statistical mechanical sum rules
were shown to play an important practical role in assessing the
quality of neural functionals obtained with simulation-based
supervised machine learning \cite{delasheras2023perspective,
  sammueller2023neural, sammueller2023whyNeural,
  sammueller2024hyperDFT, sammueller2024whyhyperDFT,
  sammueller2024pairmatching, zimmermann2024ml,
  sammueller2024attraction, buchannan2025attraction, kampa2024meta,
  sammueller2025chemicalPotential, robitschko2025mixShort}. That the
machine-learning approach entails significant potential for carrying
out efficient computational work was demonstrated in the study of
charged systems \cite{cox2020pnas, bui2024, bui2024neuralrpm,
  bui2025electromechanics, bui2025dielectrocapillarity} on the basis
of classical density functional theory~\cite{evans1979}.

Here we present the generalization of the gauge correlation framework
\cite{hermann2021noether, hermann2022topicalReview,
  hermann2022variance, hermann2022quantum, tschopp2022forceDFT,
  sammueller2023whatIsLiquid, hermann2023whatIsLiquid,
  robitschko2024any, mueller2024gauge, mueller2024whygauge,
  mueller2024dynamic} to multi-component systems in thermal
equilibrium.  The emerging species-resolved forms of the sum rules
possess similar mathematical form as in the analogous one-component
case. The species-resolved sum rules carry species labels in a
systematic way. The relative simplicity is important for the practical
applications in both theoretical and simulation work.  As a
representative model, we consider the iconic binary Lennard-Jones
fluid \cite{hansen2013}, which is a popular starting point for
investigating complex (fluid) bulk phase behaviour and associated
interfacial phenomena \cite{telodagama1983one,
  telodagama1983two, napari1999, wilding1997, wilding2002,
  schmid2001wetting, wilding2003, wilding1998, koefinger2006epl,
  koefinger2006jcp}, as addressed via classical density functional theory \cite{telodagama1983one,
  telodagama1983two, napari1999} and in simulations \cite{wilding1997, wilding2002,
  schmid2001wetting, wilding2003, wilding1998, koefinger2006epl,
  koefinger2006jcp}. A~particular symmetrical parameterization
investigated by Wilding {\it et~al.}~\cite{wilding1997,
  schmid2001wetting, wilding2002, wilding2003, wilding1998,
  koefinger2006epl, koefinger2006jcp} provides a simple case that
features only a single common lengthscale. Furthermore we consider the
popular Kob-Andersen model \cite{kob1994, kob1995} as a prototypical
asymmetrical binary mixture.  Recent work was addressed at its phase
diagram \cite{pedersen2018} and locally favoured structures
\cite{royall2017}, devitrification processes \cite{turci2019}, a
crystallization instability~\cite{ingebrigtsen2019}, many-body
correlations \cite{luo2022}, ultrastability \cite{leoni2023}, and
aging \cite{mehri2021}.

We use these two Lennard-Jones systems to exemplify our approach,
but we stay away from questions of glass formation (Kob-Andersen
model) and the topics of capillary and interfacial phase behaviour
(Wilding {\it et~al.}'s symmetrical mixture). Although our
simulation work is carried out for pairwise interparticle
potentials, the theoretical framework is general and hence also
applies to multi-body interparticle interaction potentials.

The paper is organized as follows. In Sec.~\ref{SECphaseSpaceShifting}
we present the species-resolved gauge theory, including the
description of the microscopic model Hamiltonian
(Sec.~\ref{SECmicroscopicModel}), the thermal ensemble
(Sec.~\ref{SECensemble}), the sum rules that emerge from invariance
against species-resolved phase space shifting
(Sec.~\ref{SECspeciesResolvedPhaseSpaceShifting}), and the gauge
invariance for statistical mechanical microstates
(Sec.~\ref{SECspeciesResolvedGaugeInvariance}). In
Sec.~\ref{SECforceCorrelationFunctions} we describe the sum rules for
the force-force and force gradient correlation functions in general
inhomogeneous situations
(Sec.~\ref{SECforceCorrelationsInhomogeneous}), the reduction to
species-resolved `3g-sum rules' for bulk states
(Sec.~\ref{SECthreegRules}), and exact global and local identities
(Sec.~\ref{SECglobalTwoBody}). We present the hyperforce correlation
theory for mixtures in Sec.~\ref{SEChyperForceCorrelations}, including
the general locally-resolved framework for general observables
(Sec.~\ref{SEChyperForceGeneral}), the associated global sum rules
(Sec.~\ref{SEChyperForceGlobal}), and the application to several
specific choices both within the locally-resolved
(Sec.~\ref{SEChyperForceSpecific}) and the global
(Sec.~\ref{EQspecifcGlobalSumRules}) cases.  In
Sec.~\ref{SECsimulationResults} we present our simulation results for
the bulk force-force correlation structure of the Kob-Andersen liquid
(Sec.~\ref{SECbulkStructure}) and for the confined symmetric
Lennard-Jones system (Sec.~\ref{SECconfinement}).  
In Sec.~\ref{SECconclusions} we present our conclusions and give an
outlook on possible future work.

\section{Shifting gauge transformation}
\label{SECphaseSpaceShifting}

Our treatment of multi-component mixtures is based on the statistical
mechanical invariance theory for one-component
systems~\cite{hermann2021noether, hermann2022topicalReview,
  hermann2022variance, hermann2022quantum, tschopp2022forceDFT,
  sammueller2023whatIsLiquid, hermann2023whatIsLiquid,
  robitschko2024any, mueller2024gauge, mueller2024whygauge,
  mueller2024dynamic, rotenberg2024spotted,
  miller2025physicsToday}. We give a brief account of this prior work
and refer the Reader for a discussion of the relationship to the
classical liquid state literature to Refs.~\cite{hermann2021noether,
  mueller2024whygauge}.
For homogeneous displacements Noether's theorem was shown to yield a
range of classical and novel exact sum rules for equilibrium and
nonequilibrium many-body systems~\cite{hermann2021noether}. The
application to one-dimensional systems is given in
Ref.~\cite{hermann2022topicalReview}, together with a description of
elementary statistical mechanical background. Global sum rules for the
force variance (second moment) follow from considering shifting at
second order in the displacement vector~\cite{hermann2022variance}.

Spatially inhomogeneous phase space shifting yields locally resolved
force sum rules \cite{hermann2022quantum, tschopp2022forceDFT}. At
second order in the displacement field one finds a `3g'-sum rule that
relates the pair distribution function to the force-force and
force-gradient two-body correlation functions; the latter were shown
to give deep insight into the spatial structure of liquids, networks,
and liquid crystal phases \cite{sammueller2023whatIsLiquid,
  hermann2023whatIsLiquid}. Addressing general observables of interest
\cite{robitschko2024any} yields generalized forces, which were dubbed
hyperforces inline with Hirschfelder's generalization of the standard
virial theorem \cite{hansen2013} to his hypervirial theorem
\cite{hirschfelder1960}. To clarify intent, an observable that is
subject to the treatment is referred to as a hyperobservable.
The phase space shifting transformation was identified as a gauge
transformation for equilibrium statistical mechanics
\cite{mueller2024gauge}. The shifting vector field plays the role of
the gauge function, see Ref.~\cite{mueller2024whygauge} for a
description of the analogy with classical electrodynamics. The theory
features nontrivial Lie algebra structure. A dynamical version was
presented recently \cite{mueller2024dynamic} and we refer the Reader
to Refs.~\cite{rotenberg2024spotted, miller2025physicsToday} for
popular accounts.

\bigskip

\subsection{Microscopic multi-component model}
\label{SECmicroscopicModel}
We consider systems with $M$ distinct species of particles. Each
species $\alpha=1,\ldots,M$ consists of $N_\alpha$ particles that
possess identical properties. To implement the book-keeping of the
different components, we group all particle indices~$i$ that
constitute the species $\alpha$ together into an index set~${\cal
  N}_\alpha$. Summing over all particles of species~$\alpha$ can then
be written succinctly as $\sum_{i\in {\cal N}_\alpha}$. Further
summation over all species is expressed as the sum $\sum_\alpha$,
where the summation index $\alpha$ runs over all species~$1,\ldots,
M$. An example for this mechanism is the total number of particles
$N=\sum_\alpha N_\alpha$. As a special (and admittedly extreme) case,
this labelling allows one to address all particles individually, via
setting $M=N$ and $N_\alpha=1$ for all $\alpha$. Each index set then
contains a single element, ${\cal N}_\alpha=\{\alpha\}$, and the sum
over particles of this species collapses to the single contribution
$\alpha=i$, hence effectively rendering $\alpha$ the particle index.

The microscopic model in $d$ spatial dimensions is described on the
basis of its position coordinates $\rv_1,\ldots,\rv_N\equiv \rv^N$ and
the linear momenta $\pv_1,\ldots,\pv_N\equiv \pv^N$. The Hamiltonian
$H$ contains kinetic, interparticle, and external energy contributions
according to the following standard form:
\begin{align}
  H &= \sum_\alpha \sum_{i\in{\cal N}_\alpha} \frac{\pv_i^2}{2m_\alpha} 
  + u(\rv^N) 
  + \sum_\alpha\sum_{i\in{\cal N}_\alpha} V_{\rm ext}^{(\alpha)}(\rv_i),
  \label{EQHamiltonian2}
\end{align}
where $m_\alpha$ denotes the mass of particles of species $\alpha$,
$u(\rv^N)$ is the interparticle interaction potential, and the
one-body external potential $V_{\rmext}^{(\alpha)}(\rv)$ acts on
species~$\alpha$ at position~$\rv$. That different particles of the
same species behave in the same way is encoded in the permutation
symmetry of the interparticle interaction potential $u(\rv^N)$ such
that the value of $u(\rv^N)$ remains unchanged upon interchanging the
positions of two particles of the same species. A common form of
$u(\rv^N)$ is generated by pair potentials $\phi_\aas(r)$ that act
between two particles of species~$\alpha$ and $\alpha'$ that are
separated by a center-center distance $r$. The total interparticle
potential can then be written explicitly as
$u(\rv^N)=\sum_{\alpha}\sum_{\alpha'}\sum_{i\in{\cal
    N}_\alpha}\sum'_{j\in{\cal N}_{\alpha'}}
\phi_\aas(|\rv_i-\rv_j|)/2$, where the primed sum indicates that the
case $i=j$ has been omitted, the factor $1/2$ corrects for double
counting, the species-swap symmetry
$\phi_\aas(r)=\phi_{\alpha'\alpha}(r)$ is implied, and the sums over
particle indices $\alpha$ and $\alpha'$ each run over all species
$1,\ldots,M$.
Our theoretical framework is general though and applies to
multi-body interparticle potentials, as does its one-component
version~\cite{sammueller2023whatIsLiquid,
  hermann2023whatIsLiquid}.

\subsection{Equilibrium ensemble and one-body observables}
\label{SECensemble}

The statistical mechanics of the mixture is formulated in the standard
way and we work specifically in the grand ensemble.  Formally
analogous derivations in the canonical ensemble yield, for fixed
number of particles, sum rules that are identical in form to the grand
canonical versions. We exemplify explicitly below in
Sec.~\ref{SECsimulationResults} the validity both with adaptive
Brownian dynamics, as representing the canonical ensemle, as well as
with grand canonical Monte Carlo simulations, as representing the
coupling to a particle bath.

At temperature~$T$ and species-resolved chemical potentials
$\mu_1,\ldots,\mu_M$ the grand potential~$\Omega$ and the grand
partition sum $\Xi$ are given respectively by
\begin{align}
  \Omega &= -k_BT\ln\Xi,
  \label{EQgrandPotential}\\
  \Xi &= \Tr {\rm e}^{-\beta(H-\sum_\alpha \mu_\alpha N_\alpha)},
  \label{EQgrandPartitionSum}
\end{align}
where $k_B$ denotes the Boltzmann constant and $\beta=1/(k_BT)$. The
classical trace operation in the grand ensemble is given by $\Tr
\cdot=\sum_{N_1} \ldots \sum_{N_M} (N_1!\ldots N_M!h^{dN})^{-1} \int
d\rv^N\int d\pv^N \cdot$, where the sums over particle numbers
$N_1,\ldots, N_M$ each range from $0$ to $\infty$, the symbol $h$
indicates the Planck constant, and the phase space integral is
abbreviated as $\int d\rv^N\int d\pv^N\,\cdot\,=\int d\rv_1\ldots d
\rv_N \int d\pv_1\ldots d\pv_N\,\cdot\,$. The corresponding grand
ensemble probability distribution (Gibbs measure) is
\begin{align}
  \Psi &= {\rm e}^{-\beta(H-\sum_\alpha\mu_\alpha N_\alpha)}/\Xi,
\label{EQGibbsMeasure}
\end{align}
where the normalization factor $\Xi$ is the grand partition sum
\eqref{EQgrandPartitionSum} and thermal averages can then be written
in the compact form $\langle\cdot\rangle=\Tr \cdot\, \Psi $.

We give a summary of several relevant averages that characterize the
mixture. The partial density profile $\rho_\alpha(\rv)$ of species
$\alpha$ is the average of the corresponding one-body density
``operator'' (phase space function),
$\rho_\alpha(\rv)=\langle\hat\rho_\alpha(\rv)\rangle$, where the
microscopic density observable of species $\alpha$ is given as
$\hat\rho_\alpha(\rv)=\sum_{i\in {\cal N}_\alpha}\delta(\rv-\rv_i)$,
with $\delta(\cdot)$ denoting the Dirac distribution.
Correspondingly, the mean interparticle force density that acts on
species $\alpha$ is
$\Fv_{\rmint}^{(\alpha)}(\rv)=\langle\hat\Fv_{\rmint}^{(\alpha)}(\rv)\rangle$,
where the species-resolved interparticle force density observable is
defined as
\begin{align}
  \hat\Fv_{\rmint}^{(\alpha)}(\rv)=-\sum_{i\in{\cal
      N}_\alpha} \delta(\rv-\rv_i)\nabla_i u(\rv^N),
\end{align}
where $\nabla_i$ denotes the derivative with respect to $\rv_i$.
Similarly, the species-resolved average kinetic stress is
$\taub_\alpha(\rv)=\langle\hat\taub_\alpha(\rv)\rangle$, with the
kinetic stress observable being defined as
\begin{align}
  \hat\taub_\alpha(\rv)=-\sum_{i\in{\cal
      N}_\alpha}\delta(\rv-\rv_i)\frac{\pv_i\pv_i}{m_\alpha},
  \label{EQtaubDefinition}
\end{align}
where $\pv_i\pv_i$ indicates the dyadic product of the momentum of
particle~$i$ with itself. These observables can be combined into a
species-resolved total force operator (phase space function), given as
\begin{align}
  \hatFva(\rv)=\nabla \cdot
  \hat\taub_\alpha(\rv) + \hat \Fv_{U}^{(\alpha)}(\rv),
  \label{EQforceDensityOperatorMixture}
\end{align}
where $\nabla$ denotes the derivative with respect to $\rv$, and the
potential force density observable for species $\alpha$ is
$\hat\Fv_{U}^{(\alpha)}(\rv)= \hat\Fv_{\rmint}^{(\alpha)}(\rv) -
\hat\rho_\alpha(\rv)\nabla V_{\rmext}^{(\alpha)}(\rv)$.  The averaged
total force density acting on species $\alpha$ then follows as the
average $\Fva(\rv)=\langle \hatFva(\rv)\rangle$. The species-resolved
potential-only force density is $\Fv_{U}^{(\alpha)}(\rv)=\langle
\hat\Fv_{U}^{(\alpha)}(\rv)\rangle$.
The thermal average of the divergence of the kinetic stress
\eqref{EQtaubDefinition} simplifies as $\nabla\cdot \langle \hat\taub_\alpha(\rv) \rangle = -k_BT \nabla\rho_\alpha(\rv)$, as follows straightforwardly from calculating the second
moments of the Maxwell distribution.

\subsection{Species-resolved phase space shifting}
\label{SECspeciesResolvedPhaseSpaceShifting}

In order to identify the thermal invariance of the mixture, we
introduce shifting fields that are unique for each component~$\alpha$
of the mixture, in generalization of the local shifting transformation
for pure systems \cite{tschopp2022forceDFT, hermann2022quantum,
  sammueller2023whatIsLiquid, hermann2023whatIsLiquid,
  robitschko2024any, mueller2024gauge,
  mueller2024whygauge}. Specifically, the species-resolved
transformations are:
\begin{align}
    \rv_i &\to \rv_i + \eps_\alpha(\rv_i),
  \label{EQmixtureCanonicalTransformationCoordinates}\\
  \pv_i &\to [\unity + \nabla_i\eps_\alpha(\rv_i)]^{-1}\cdot\pv_i,
  \label{EQmixtureCanonicalTransformationMomenta}
\end{align}
where particle $i$ is of type $\alpha$ such that $i\in{\cal N}_\alpha$
and $\eps_\alpha(\rv)$ is the $d$-dimensional vector field that
displaces particles of component~$\alpha$. In
\eqr{EQmixtureCanonicalTransformationMomenta} the symbol $\unity$
denotes the $d\times d$-unit matrix, $\nabla_i$ is the derivative with
respect to $\rv_i$, and matrix inversion is indicated by the
superscript $-1$. The transformations
\eqref{EQmixtureCanonicalTransformationCoordinates} and
\eqref{EQmixtureCanonicalTransformationMomenta} retain the canonical
properties of the one-component version \cite{tschopp2022forceDFT,
  hermann2022quantum, sammueller2023whatIsLiquid,
  hermann2023whatIsLiquid}, as the latter already acts merely
individually on the position and the momentum of each particle~$i$.

The generalized transformation
\eqref{EQmixtureCanonicalTransformationCoordinates} and
\eqref{EQmixtureCanonicalTransformationMomenta} allows one to address
the individual species separately. The strategy of the subsequent
argumentation carries over straightforwardly from the one-component
case \cite{tschopp2022forceDFT, hermann2022quantum,
  sammueller2023whatIsLiquid, hermann2023whatIsLiquid}, as we lay out
in the following. The invariance of the grand potential implies
$\Omega[\{\eps_{\alpha'}\}]=\Omega$, where on the right hand side
$\Omega$ with no argument is, as before, the grand potential
\eqref{EQgrandPotential} expressed in the original phase space
variables. On the left hand side $\{\eps_{\alpha'}\}$ indicates the
set of all displacement fields $\{\eps_1(\rv),\ldots,\eps_M(\rv)\}$,
which are used to transform the phase space variables.

The general invariance of the grand potential holds for every order upon expansion in $\{\eps_{\alpha'}(\rv)\}$.
We consider invariance at first order in the displacement fields and
follow the arguments for the one-component case
\cite{tschopp2022forceDFT, hermann2022quantum,
  sammueller2023whatIsLiquid, hermann2023whatIsLiquid}, which allow
one to conclude that $\delta \Omega[\{\eps_{\alpha'}\}]/\delta
\eps_\alpha(\rv)=0$. Explicitly carrying out the functional derivative
gives the following exact species-resolved force density balance:
\begin{align}
  -k_BT\nabla \rho_\alpha(\rv) + \Fv_{\rmint}^{(\alpha)}(\rv)
  -\rho_\alpha(\rv) \nabla V_{\rmext}^{(\alpha)}(\rv) &= 0.
  \label{EQforceDensityBalanceMixture}
\end{align}
The derivation of \eqr{EQforceDensityBalanceMixture} rests on the
following operator identity, which is obtained from expressing the
Hamiltonian \eqref{EQHamiltonian2} in the new coordinates such that it
carries an apparent functional dependence on the set of shifting
fields,
\begin{align}
  -\frac{\delta H[\{\eps_{\alpha'}\}]}{\delta\eps_\alpha(\rv)}
  \Big|_{\{\eps_{\alpha'}=0\}}
  &=  \hatFva(\rv).
  \label{EQdeltaHdeltaEpsMixture}
\end{align}
The right hand side of \eqr{EQdeltaHdeltaEpsMixture} consists of the
kinetic, interparticle, and external force densities described in
Sec.~\ref{SECensemble}. The thermal average of $\hat\Fv_\alpha(\rv)$
is generated via applying the functional derivative to the grand
potential~$\Omega$, see its definition \eqref{EQgrandPotential}, and
the arguments below \eqr{EQforceDensityOperatorMixture}. These
steps lead to \eqr{EQforceDensityBalanceMixture}, which can be written
in more compact form as
\begin{align}
  \Fv_\alpha(\rv) &= 0.
  \label{EQforceDensityBalanceMixtureCompact}
\end{align}
For details about specific steps we refer the Reader to the
description of the one-component case in
Ref.~\cite{hermann2023whatIsLiquid}.

\subsection{Statistical mechanical gauge invariance}
\label{SECspeciesResolvedGaugeInvariance}

For the case of one-component systems, $M=1$, the phase space variable
transformation \eqref{EQmixtureCanonicalTransformationCoordinates} and
\eqref{EQmixtureCanonicalTransformationMomenta} was shown to
constitute a gauge transformation of the statistical mechanical
microstates \cite{mueller2024gauge, mueller2024whygauge}. Even though
the microstates are transformed, any equilibrium average remains
invariant under the transformation. In particular the phase space
shifting is shown to be closely associated with a specific
differential operator structure on phase space. These ``shifting
differential operators'' apply to general phase space functions and
they perform a role analogous to that of the explicit coordinate
transformation \eqref{EQmixtureCanonicalTransformationCoordinates} and
\eqref{EQmixtureCanonicalTransformationMomenta}. Here we generalize
the statistical mechanical gauge invariance concept to mixtures and
thus define the following species- and position-resolved phase space
differential operators:
\begin{align}
  \bsig_\alpha(\rv) &= \sum_{i \in \calN_\alpha}[
  \delta(\rv-\rv_i)\nabla_i
  +\pv_i  \nabla \delta(\rv-\rv_i) \cdot \nabla_{\pv_i}
  ],
  \label{EQbsig}
\end{align}
where $\nabla_{\pv_i}$ denotes the derivative with respect to $\pv_i$
and $\pv_i\nabla$ is a dyadic product \cite{mueller2024gauge,
  mueller2024whygauge}. The crucial difference to the one-component
version $\bsig(\rv)$ \cite{mueller2024gauge, mueller2024whygauge} is
the mere restriction of particle summation from a sum over all
particles to $\sum_{i\in{\cal N}_\alpha}$ in \eqr{EQbsig}.

The operators \eqref{EQbsig} are anti-self-adjoint on phase space and
they satisfy nontrivial commutator structure, as respectively
expressed by
\begin{align}
  \bsig_\alpha^\dagger(\rv) &= -\bsig_\alpha(\rv),
  \label{EQsigmaAntiSelfAdjoint}\\
  [\bsig_\alpha(\rv), \bsig_\alphas(\rv')] &= 
  \delta_{\aas} \bsig_\alpha(\rv')[\nabla \delta(\rv-\rv')]
  \notag \\ & \quad
  +\delta_{\aas} [\nabla\delta(\rv-\rv')]\bsig_\alpha(\rv),
  \label{EQsigmaCommutator}
\end{align}
where $\delta_{\aas}$ denotes the Kronecker symbol and the dagger
indicates the adjoint, which for an operator ${\cal O}$ and two
general phase space functions $\hat A(\rv^N, \pv^N)$ and $\hat
B(\rv^N, \pv^N)$ is defined in the standard way via $\int d\rv^N
d\pv^N \hat A {\cal O} \hat B = \int d\rv^N d\pv^N \hat B {\cal
  O}^\dagger \hat A$.

The localized shifting operators \eqref{EQbsig} can be combined
together with their respective shifting fields $\eps_\alpha(\rv)$,
which play the role of gauge functions, to define integrated shifting
operators
\begin{align}
  \Sigma[\{\eps_\alpha\}] &= \sum_\alpha
  \int d\rv \eps_\alpha(\rv) \cdot \bsig_\alpha(\rv),
\end{align}
where on the left hand side the bracketed argument indicates the
functional dependence on the set of shifting fields $\{\eps_1(\rv),
\ldots, \eps_M(\rv)\}$. A given phase space function $\hat A(\rv^N,
\pv^N)$ is then affected by the transformation to lowest order in the
shifting fields and their spatial gradients as
\begin{align}
  \hat A(\tilde\rv^N,\tilde\pv^N) &=
  \hat A(\rv^N, \pv^N) + \Sigma[\{\eps_\alpha\}]
  \hat A(\rv^N, \pv^N),
\end{align}
where the tilde indicates the new phase space
variables~\eqref{EQmixtureCanonicalTransformationCoordinates} and
\eqref{EQmixtureCanonicalTransformationMomenta}.
The operators $\Sigma[\{\eps_\alpha\}]$ satisfy nontrivial Lie algebra
structure, such that the commutator is $[\Sigma[\{\eps_\alpha\}],
  \Sigma[\{\eps'_\alpha\}]] = \Sigma[\{\eps''_\alpha(\rv)\}]$, where
the difference shifting field is given by $\eps''_\alpha(\rv) =
\eps_\alpha(\rv)\cdot[\nabla\eps_\alpha'(\rv)] -
\eps'_\alpha(\rv)\cdot[\nabla\eps_\alpha(\rv)]$.  Hence the integrated
shifting operators $\Sigma[\{\eps_\alpha\}]$ continue to satisfy the
Lie algebra structure described in Ref.~\cite{mueller2024gauge},
including the Jacobi identity.

Applying the localized shifting operators \eqref{EQbsig} to a given
phase space function $\hat A(\rv^N, \pv^N)$ is identical to carrying
out the following functional differentiation operations at first order:
\begin{align}
  \bsig_\alpha(\rv)\hat A
  &= \frac{\delta \hat A(\tilde \rv^N, \tilde \pv^N)}{\delta \eps_\alpha(\rv)}
  \Big|_{\{\eps_\alpha=0\}},
  \label{EQbsigAtFirstOrder}
\end{align}
and at second order:
\begin{align}
  \bsig_\alpha(\rv) \bsig_\alphas(\rv') \hat A 
  &= \frac{\delta^2 \hat A(\tilde\rv^N, \tilde\pv^N)}
       {\delta\eps_\alpha(\rv)\delta\eps_\alphas(\rv')}
       \Big|_{\{\eps_\alpha=0\}}
       \notag\\&\quad
       +\delta_\aas[\nabla\delta(\rv-\rv')]
       \bsig_\alpha(\rv) \hat A(\rv^N,\pv^N).
       \label{EQseventeen}
\end{align}
The (phase space) arguments of $\hat A(\rv^N, \pv^N)$ are suppressed
on the above left hand sides for brevity. As indicated in the notation
on the right hand sides, all partial shifting fields
$\eps_\alpha(\rv)$ are set to zero after the functional derivatives
have been taken. Equations~\eqref{EQbsigAtFirstOrder} and
  \eqref{EQseventeen} follow from argumentation that is analogous to
  the corresponding one-component versions, see
  Ref.~\cite{mueller2024whygauge}.
In particular, when choosing the Hamiltonian as the
hyperobservable, $\hat A=H$ in \eqr{EQbsigAtFirstOrder}, then
comparison to \eqr{EQdeltaHdeltaEpsMixture} yields the
species-resolved force density observable as $\hat
\Fv_\alpha(\rv)=-\bsig_\alpha(\rv) H$, as is consistent with
applying the explicit form \eqref{EQbsig} to $-H$.

When applied to a specific observable $\hat A$ the species-resolved
hyperforce density is $\hat\Sv_{A}^{(\alpha)}(\rv) =
\bsig_\alpha(\rv)\hat A$, see Eqs.~\eqref{EQbsig} and
\eqref{EQbsigAtFirstOrder}. Explicitly the resulting phase space form
of the species-resolved hyperforce density is
\begin{align}
  \hat\Sv_{A}^{(\alpha)}(\rv) &=
  \sum_{i \in{\cal N}_\alpha} \big[ \delta(\rv-\rv_i)\nabla_i \hat A
    + \pv_i   \nabla \delta(\rv-\rv_i) \cdot \nabla_{\pv_i} \hat A\big].
  \label{EQSvAdefinition}
\end{align}
For completeness, when applying \eqr{EQSvAdefinition} to the
Hamiltonian, i.e.\ upon choosing $\hat A=H$, one obtains the
(negative) force density observable, $\hat\Sv_{A=H}^{(\alpha)}(\rv)
= -\hat\Fv_\alpha(\rv)$.

From the commutator relationship \eqref{EQsigmaCommutator} the
following Lie sum rules are obtained upon phase space averaging:
\begin{align}
  &  \langle \hat\Sv_{A}^{(\alpha)}(\rv)\beta\hat\Fv_\alphas(\rv')\rangle
  -\langle \beta \hat\Fv_\alpha(\rv)\hat\Sv_{A}^{(\alphas)}(\rv')\rangle
  \notag\\&  \qquad=
  \delta_\aas\big\{ 
  \Sv_{A}^{(\alpha)}(\rv') [\nabla \delta(\rv-\rv')]
  + [\nabla\delta(\rv-\rv')]\Sv_{A}^{(\alpha)}(\rv)
  \big\},
  \label{EQsumRuleLieMixtures}
\end{align}
where the argumentation is analogous to the one-component
treatment \cite{mueller2024whygauge}.
As a special case, when the right hand side of
\eqref{EQsumRuleLieMixtures} vanishes, we have:
\begin{align}
  \langle \hat\Sv_{A}^{(\alpha)}(\rv)\hat\Fv_\alphas(\rv')\rangle
  - \langle \hat \Fv_\alpha(\rv)\hat\Sv_{A}^{(\alphas)}(\rv') \rangle
  &= 0,
\end{align}
which holds true provided that $\alpha\neq\alpha'$ or $\rv\neq\rv'$.

We next describe several concrete consequences of the gauge invariance
for the correlation structure of soft matter mixtures.

\section{Force correlation functions}
\label{SECforceCorrelationFunctions}

\subsection{Inhomogeneous partial pair force correlations}
\label{SECforceCorrelationsInhomogeneous}

Addressing second-order phase space shifting, we consider the
functional Hessian with respect to the shifting fields, i.e.\ the
species-resolved second derivatives $\delta^2
\Omega[\{\eps_{\alpha''}\}]/\delta\eps_\alpha(\rv)\delta
\eps_{\alpha'}(\rv')=0$. One finds upon carrying out the explicit
calculation on the basis of \eqr{EQdeltaHdeltaEpsMixture} the
following result:
\begin{align}
  \beta \langle 
  \hatFva(\rv) \hatFvas(\rv')
  \rangle &=
  \Big\langle
  \frac{\delta^2 H[\{\eps_{\alpha''}\}]}
       {\delta\eps_\alpha(\rv)\delta\eps_{\alpha'}(\rv')}
       \Big\rangle
       \Big|_{\{\eps_{\alpha''}=0\}},
       \label{EQsumRule2genericMixture}
\end{align}
where again the shifting fields are set to zero after the functional
derivatives of the Hamiltonian have been taken. We recall that
$\hatFva(\rv)$ indicates the species- and position-resolved total
force density observable \eqref{EQforceDensityOperatorMixture}, which
includes the potential forces and the divergence of the kinetic stress
contribution. The sum rule \eqref{EQsumRule2genericMixture}
generalizes the corresponding one-component identity
\cite{sammueller2023whatIsLiquid, hermann2023whatIsLiquid}.  It is
useful to split off the potential forces in the sum rule
\eqref{EQsumRule2genericMixture} and to furthermore also discriminate
between self and distinct cases according to whether the same or two
different particles contribute to the occurring double sums. In the
derivation one can make use of \eqr{EQseventeen}, setting $\hat A= H$
therein, and we refer to Ref.~\cite{hermann2023whatIsLiquid} for
further details on the corresponding reasoning.

The resulting position-dependent two-body distinct sum rule has the
following species-resolved form:
\begin{align}
&  \big\langle
  \beta \hat\Fv_{U}^{(\alpha)}(\rv)
  \beta \hat\Fv_{U}^{(\alpha')}(\rv')
  \big\rangle_{\rm dist}
  = \nabla\nabla' \rho_{2}^{(\aas)}(\rv,\rv')
  \notag\\
  &\quad
  +\Big\langle
  \sum_{i\in {\cal N}_\alpha} \; \sideset{}{'}\sum_{j\in{\cal N}_{\alpha'}}
  \delta(\rv-\rv_i)\delta(\rv'-\rv_j)
  \nabla_i\nabla_j\beta u(\rv^N)
  \Big\rangle,
  \label{EQsumRuleFeFeDistinctMixture}
\end{align}
where the (distinct) two-body density is defined as the thermal
average $\rho_{2}^{(\aas)}(\rv,\rv')=\langle\hat\rho_\alpha(\rv)
\hat\rho_{\alpha'}(\rv')\rangle_{\rm dist}= \langle\sum_{i\in{\cal
    N}_\alpha}\sum'_{j\in{\cal
    N}_{\alpha'}}\delta(\rv-\rv_i)\delta(\rv'-\rv_j)\rangle$.  We
recall that the primed sum indicates the restriction $j\neq i$, which
only plays a role for the case of identical species,
$\alpha=\alpha'$. The distinct average on the left hand side of
\eqr{EQsumRuleFeFeDistinctMixture} correspondingly excludes the case
$i=j$ in the occurring double sums over particles when writing out the
two potential force operators.

The corresponding self part of the sum rule follows from considering
double occurrences of the same particle in
\eqr{EQsumRule2genericMixture}, which leads to the following exact
identity:
\begin{align}
  &\big\langle \beta \hat\Fv_{U}^{(\alpha)}(\rv) 
    \beta \hat\Fv_{U}^{(\alpha)}(\rv)\big\rangle_{\rm self}
  = \nabla\nabla\rho_\alpha(\rv)\notag\\  &\quad  
  +\avg{\sum_{i\in{\cal N}_\alpha}\delta(\rv-\rv_i)\nabla_i\nabla_i\beta u(\rv^N)}
  +\rho_\alpha(\rv)\nabla\nabla \beta V_{\rmext}^{(\alpha)}(\rv).
  \label{EQsumRuleFeSelfMixture}
\end{align}
The self part on the left hand side of \eqr{EQsumRuleFeSelfMixture}
involves only the case of the same particle occurring twice in the
double sum, such that this term constitutes the following dyadic
product $\langle \beta^2 \sum_{i\in {\cal N}_\alpha}
\delta(\rv-\rv_i)[\nabla_i \hat U][\nabla_i \hat U]\rangle$, where
$-\nabla_i\hat U(\rv^N)=-\nabla_i [u(\rv^N)+
  V_{\rmext}^{(\alpha)}(\rv_i)]$ is the potential force that acts on
particle $i$.

\subsection{Bulk fluid 3g-sum rule for mixtures}
\label{SECthreegRules}

We can simplify the above general two-body framework by resolving only
the dependence on the relative distance between the two positions
$\rv$ and $\rv'$. The standard partial (species-labelled) pair
distribution function is given as
\begin{align}
  g_\aas(r) = \frac{\rho_2^{(\aas)}(\rv,\rv')}{\rho_\alpha^b\rho_\as^b},
\end{align}
where $\rho_\alpha^b$ indicates the bulk density of species $\alpha$,
the distance is $r=|\rv-\rv'|$, and we assume homogeneous and
  isotropic fluid states. Analogously, the partial bulk force-force
and force-gradient pair correlation functions are defined respectively
as
\begin{align}
  & {\sf g}_{\ff}^{(\aas)}(r) =
  \frac{\beta^2}{\rho_\alpha^b\rho_\as^b}
  \langle \hat\Fv_{\rmint}^{(\alpha)}(\rv)
  \hat\Fv_{\rmint}^{(\alpha')}(\rv') \rangle_{\rm dist},\\
  & {\sf g}_{\gradf}^{(\aas)}(r) = \notag\\ &\quad
  -\frac{\beta}{\rho_\alpha^b\rho_\as^b}
  \avg{ \sum_{i\in{\cal N}_\alpha} 
    \sideset{}{'} \sum_{j\in{\cal N}_\as}
    \delta(\rv-\rv_i)\delta(\rv'-\rv_j) \nabla_i \nabla_j u(\rv^N)}.
\end{align}

\begin{figure*}[tb]
  \includegraphics[page=1,width=.8\textwidth]{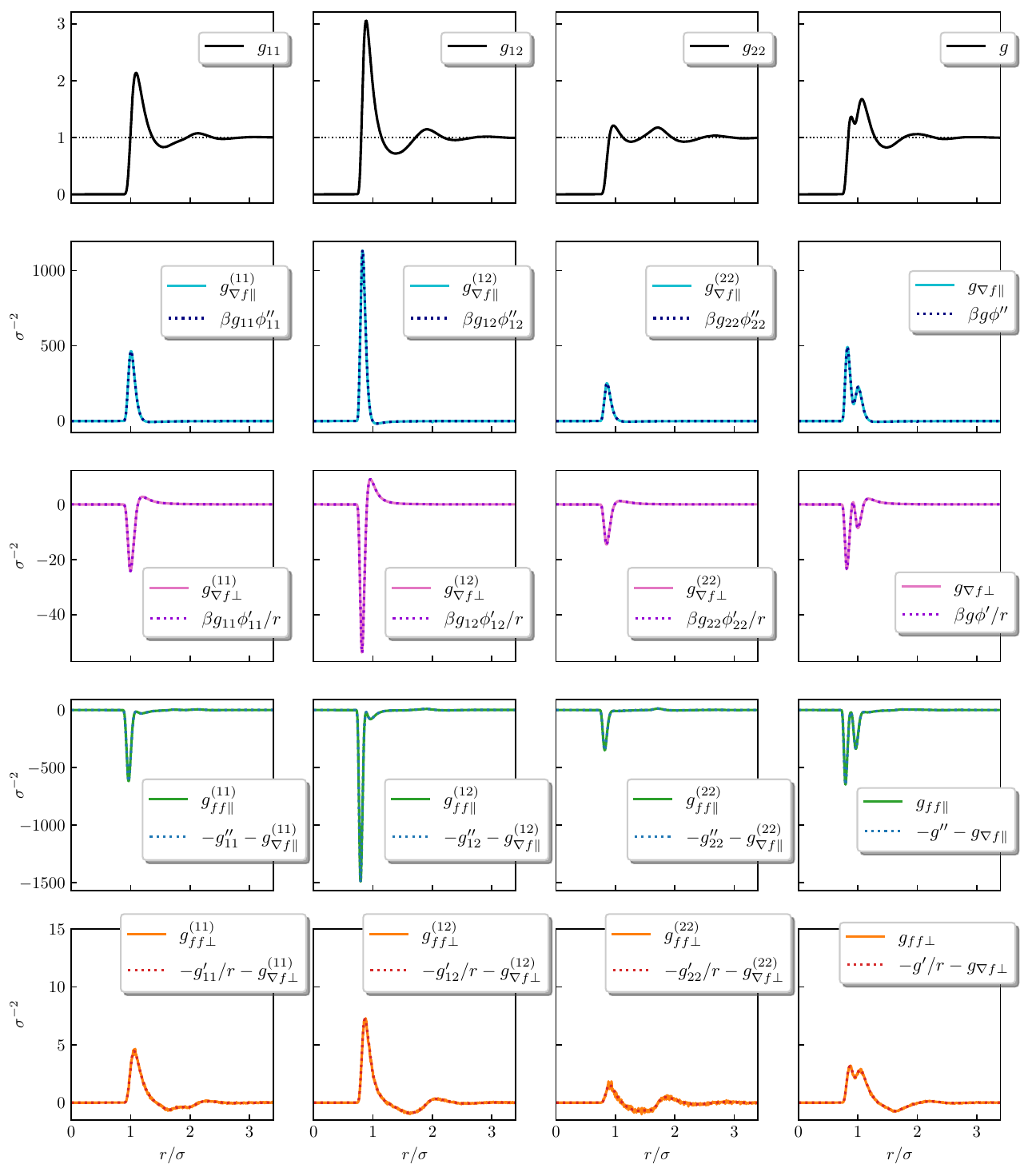}
  \caption{Two-body gauge correlation functions of the Kob-Andersen
    liquid at reduced temperature $k_BT/\epsilon=1.1$ and scaled
    partial bulk densities $\rho_1^b\sigma^3=0.591$ and
    $\rho_2^b\sigma^3=0.253$.
    Results are shown as a function of the scaled interparticle
    distance $r/\sigma$ for species $\aas=11$ (first column), 12
    (second column), 22 (third column), and for the agglomerated
    quantities (fourth column).
    Top row: partial pair distribution functions $g_{\aas}(r)$.
    Second row: the parallel component of the force-gradient
    correlation function $g_{\nabla f\parallel}^{(\aas)}(r)$ agrees
    numerically with $g_\aas(r)\beta \phi_\aas''(r)$,
    cf.~\eqr{EQforceGradientViaPairPotential}.
    Third row: the corresponding perpendicular component $g_{\nabla
      f\perp}^{(\aas)}(r)$ agrees numerically with
    $g_\aas(r)\beta\phi_\aas'(r)/r$,
    cf.~\eqr{EQforceGradientViaPairPotential}.
    Fourth row: the parallel component of the force-force correlation
    function $g_{f\!f\parallel}^{(\aas)}(r)$ agrees numerically with
    $-g_\aas''(r) - g_{\gradf\parallel}^{(\aas)}(r)$, as expected from
    the sum rules \eqref{EQ3gParallel} (first to third panel) and
    \eqref{EQgIdentityAgglomeratedParallel} (last panel).
    Bottom row: the corresponding perpendicular component
    $g_{f\!f\perp}^{(\aas)}(r)$ agrees numerically with $-g_\aas'(r)/r
    - g_{\gradf\perp}^{(\aas)}(r)$, as expected from sum
    rules~\eqref{EQ3gPerpendicular} (first to third panels) and
    \eqref{EQgIdentityAgglomeratedPerpendicular} (last panel).  }
\label{FIGgofr}
\end{figure*}

For a bulk mixture, where $V_{\rmext}^{(\alpha)}(\rv)=0$ for all $\alpha$,
we can simplify the inhomogeneous force correlation sum rule
\eqr{EQsumRuleFeFeDistinctMixture} to obtain the following homogeneous
form:
\begin{align}
  \nabla\nabla g_{\aas}(r) +
  {\sf g}_{\gradf}^{(\aas)}(r)
  +{\sf g}_{\ff}^{(\aas)}(r) &= 0,
  \label{EQgIdentityMix}
\end{align}
which reduces to the one-component 3g-sum rule $\nabla\nabla g(r) +
{\sf g}_\gradf(r) + {\sf g}_\ff(r)=0$ in the case of single species
\cite{sammueller2023whatIsLiquid, hermann2023whatIsLiquid}. We have
generalized the three pair correlation functions via restriction of
the sums over all particles to sums over the appropriate index
sets. Furthermore we have replaced the normalization factor $\rho_b^2$
by $\rho^b_\alpha\rho^b_{\alpha'}$, where the bulk number density of
species $\alpha$ is $\rho^b_\alpha=\langle N_\alpha \rangle/V$; we
recall that $N_\alpha$ is the number of particles of species $\alpha$,
and $V$ indicates the system volume.

As in the one-component case \cite{sammueller2023whatIsLiquid,
  hermann2023whatIsLiquid}, the two nontrivial spatial components of
the tensorial identity \eqref{EQgIdentityMix} are parallel
($\parallel$) and perpendicular ($\perp$) to the distance vector
between the two particles and hence
\begin{align}
  g''_{\aas}(r)
  + g_{\gradf\parallel}^{(\aas)}(r)
  + g_{\ff\parallel}^{(\aas)}(r) &= 0,
  \label{EQ3gParallel}\\
  \frac{g'_{\aas}(r)}{r}
  + g_{\gradf\perp}^{(\aas)}(r)
  + g_{\ff\perp}^{(\aas)}(r) &= 0,
  \label{EQ3gPerpendicular}
\end{align}
where \eqr{EQ3gPerpendicular} holds for systems with spatial
dimensionality $d\geq 2$. The primed functions in
Eqs.~\eqref{EQ3gParallel} and \eqref{EQ3gPerpendicular} are
derivatives by the argument, such that $g'_\aas(r)$ and $g''_\aas(r)$
are respectively the first and second derivatives of the partial pair
distribution function $g_\aas(r)$ with respect to distance~$r$.

The above framework is general and applies to many-body
interparticle interaction potentials $u(\rv^N$). For fluid mixtures
in which the particles interact mutually solely via pair potentials
$\phi_\aas(r)$, we have
\begin{align}
  {\sf g}_{\gradf}^{(\aas)}(r) &=
  g_\aas(r)\beta\nabla\nabla\phi_\aas(r),
  \label{EQforceGradientViaPairPotential}
\end{align}
where $\nabla\nabla\phi_\aas(r)$ possesses one nontrivial parallel
component, $\phi_\aas''(r)$, and two identical perpendicular
components, $\phi_\aas'(r)/r$.  As a consequence, for such pairwise
interacting mixtures we can express Eqs.~\eqref{EQ3gParallel} and
\eqref{EQ3gPerpendicular} in the respective forms:
\begin{align}
  g_\aas''(r) + g_\aas(r) \beta \phi''_\aas(r)
  + g_{\ff\parallel}^{(\aas)}(r) &= 0,
  \label{EQthreegSumRulePairParallel}
  \\
  \frac{g_\aas'(r)}{r} + g_\aas(r)\frac{\beta \phi'_\aas(r)}{r}
  +g_{\ff\perp}^{(\aas)}(r)
  &= 0.
  \label{EQthreegSumRulePairPerp}
\end{align}
All sum rules reduce to their one-component versions
\cite{hermann2023whatIsLiquid} in the limit of a single
component. This also applies when considering `agglomerated'
correlation functions that ignore the species labelling, as we
demonstrate in the following. For details about the radial
dependences and the occurrences of first and second radial
derivatives in Eqs.~\eqref{EQthreegSumRulePairParallel} and
\eqref{EQthreegSumRulePairPerp} we refer the Reader to
Ref.~\cite{sammueller2023whatIsLiquid}.

We introduce species-resolved concentration variables $c_\alpha =
\rho_\alpha^b/\rho^b$ where the total bulk density is $\rho^b =
\sum_\alpha\rho_\alpha^b$.  Then summing over species yields the
following `colour-blind' or 'species-agnostic' versions as linear
combinations of the species-resolved correlation functions.  We obtain
the agglomerated pair distribution function $g(r) =
\sum_{\alpha,\alpha'} c_\alpha c_{\alpha'}g_{\alpha\alpha'}(r)$,
the force-gradient correlation function ${\sf g}_\gradf(r) =
\sum_{\alpha,\alpha'} c_\alpha c_{\alpha'}{\sf
  g}_\gradf^{(\alpha\alpha')}(r)$, which for pairwise interactions can
be written as ${\sf g}_\gradf^{(\alpha\alpha')}(r) =
\sum_{\alpha\alpha'}c_\alpha c_{\alpha'} g_{\alpha\alpha'}(r)
\nabla\nabla\beta\phi_{\alpha\alpha}(r)$,
and the force-force correlation function ${\sf g}_\ff(r) =
\sum_{\alpha,\alpha'} c_\alpha c_{\alpha'}{\sf
  g}_\ff^{(\alpha\alpha')}(r)$.
Then the 3g-sum rule is obtained in the `agglomerated' version:
\begin{align}
  \nabla\nabla g(r) +
  {\sf g}_{\gradf}(r)
  +{\sf g}_{\ff}(r) &= 0,
  \label{EQgIdentityAgglomerated}
\end{align}
which is formally identical to the single-component sum rule
\cite{sammueller2023whatIsLiquid, hermann2023whatIsLiquid}.  The two
relevant spatial tensor components are parallel and transversal to the
interparticle distance vector and they satisfy respectively the
following sum rules:
\begin{align}
  g''(r)
  + g_{\gradf\parallel}(r)
  + g_{\ff\parallel}(r) &= 0,
    \label{EQgIdentityAgglomeratedParallel}\\
  \frac{g'(r)}{r}
  + g_{\gradf\perp}(r)
  + g_{\ff\perp}(r) &= 0.
  \label{EQgIdentityAgglomeratedPerpendicular}
\end{align}
We present results from simulation work in
Sec.~\ref{SECsimulationResults} to demonstrate the accessibility of
all partial two-body gauge correlation functions together with tests
of the relevant sum rules that these satisfy.

\subsection{Global and local two-body sum rules}
\label{SECglobalTwoBody}

We have so far treated position-resolved cases where the spatial
dependence is retained.  A global second order sum rule is obtained
from either considering spatially constant displacements
$\eps_\alpha(\rv)=\eps_\alpha=\rm const$ or alternatively integrating
the spatially resolved identity \eqref{EQsumRule2genericMixture} over
both position variables and summing over both species. The result of
both routes of derivation is the same and it is given by
\begin{align}
  & \sum_{\aas}  \int d\rv d\rv' H_{2}^{(\aas)}(\rv,\rv') 
  \nabla V_{\rmext}^{(\alpha)}(\rv) \nabla' V_{\rmext}^{(\alpha')}(\rv')
  \notag\\&\quad
  = k_BT 
  \sum_\alpha
  \int d\rv \rho_\alpha(\rv) \nabla\nabla V_{\rmext}^{(\alpha)}(\rv).
  \label{EQsumRuleGlobalSecondOrderMixture}
\end{align}
Here the correlation function of density fluctuations
\cite{hansen2013, evans1979, schmidt2022rmp} has the common form
$H_{2}^{(\aas)}(\rv,\rv')=
\cov(\hat\rho_\alpha(\rv),\hat\rho_{\alpha'}(\rv'))$, where the
covariance of two operators $\hat A$ and $\hat B$ is defined in the
standard way as $\cov(\hat A, \hat B)=\langle\hat A\hat
B\rangle-\langle \hat A \rangle \langle \hat B \rangle$. Equation
\eqref{EQsumRuleGlobalSecondOrderMixture} is analogous to the
corresponding one-component identity of
Ref.~\cite{hermann2022variance}, which itself is recovered for the
case of a single species, $M=\alpha=\alpha'=1$.

A species-resolved version of \eqr{EQsumRuleGlobalSecondOrderMixture},
which involves also interparticle contributions, is given by
\begin{align}
  \beta {\rm cov}(\hat \Fv_U^{\glob(\alpha)}, \hat \Fv_U^{\glob(\alphas)}) 
  &= \Big\langle
  \sum_{i\in{\cal N}_\alpha}\sum_{j\in{\cal N}_\alphas} \nabla_i
  \nabla_j u(\rv^N) \Big\rangle
  \notag\\&\qquad
  + \delta_\aas \int d\rv \rho_\alpha(\rv)
  \nabla\nabla V_\rmext^{(\alphas)}(\rv),
  \label{EQsumRuleGlobalSecondOrderMixtureSpeciesResolved}
\end{align}  
where the species-resolved global force operator is
\begin{align}
  \hat\Fv_U^{\glob(\alpha)} = \int d\rv \hat\Fv_U^{(\alpha)}(\rv) =
  -\sum_{i\in{\cal N}_\alpha} \nabla_i \hat U.
  \label{EQglobalPotentialForceOperator}
\end{align}
Summing over both species labels $\alpha$ and $\alphas$ in
\eqr{EQsumRuleGlobalSecondOrderMixtureSpeciesResolved} and observing
that $\sum_\alpha \sum_{i\in{\cal N}_\alpha} \nabla_i u(\rv^N)=0$
recovers \eqr{EQsumRuleGlobalSecondOrderMixture}.

Furthermore the density-force correlation sum rule
\cite{hermann2023whatIsLiquid} for the case of mixtures is
\begin{align}
  \big\langle \beta\hatFva(\rv) \hat\rho_{\alpha'}(\rv') \big\rangle
  &= \delta_\aas\nabla' \rho_{2,\rmself}^{(\alpha)}(\rv,\rv'),
  \label{EQsumRuleDensityForceMixture}
\end{align}
where the self two-body density distribution is defined as
$\rho_{2,\rmself}^{(\alpha)}(\rv,\rv')=\langle \sum_{i\in{\cal
    N}_\alpha} \delta(\rv-\rv_i) \delta(\rv'-\rv_i) \rangle$ and the
right hand side of \eqr{EQsumRuleDensityForceMixture} can
alternatively be written as $\delta_{\alpha\alpha'} \rho(\rv) \nabla'
\delta(\rv-\rv')$.
The derivation of \eqr{EQsumRuleDensityForceMixture} can be based on
the mixed second order invariance $0=\delta^2 \Omega/[\delta
  \eps_\alpha(\rv)\delta V_{\rmext}^{(\alpha')}(\rv')] = \delta
\rho_{\alpha'}(\rv')/\delta \eps_\alpha(\rv) = -\delta
\Fva(\rv)/\delta V_{\rmext}^{(\alpha')}(\rv')$.  Here the two
alternative resulting expressions are obtained from exchanging the
order of the two functional derivatives, re-writing via using that
$\rho_{\alpha'}(\rv')=\delta\Omega/\delta
V_{\rmext}^{(\alpha')}(\rv')$ and $\Fva(\rv)=-\delta\Omega/\delta
\eps_\alpha(\rv)$, and setting the species-resolved displacement
fields to zero.

\begin{figure*}[tb]
  \includegraphics[page=1,width=0.8\textwidth]{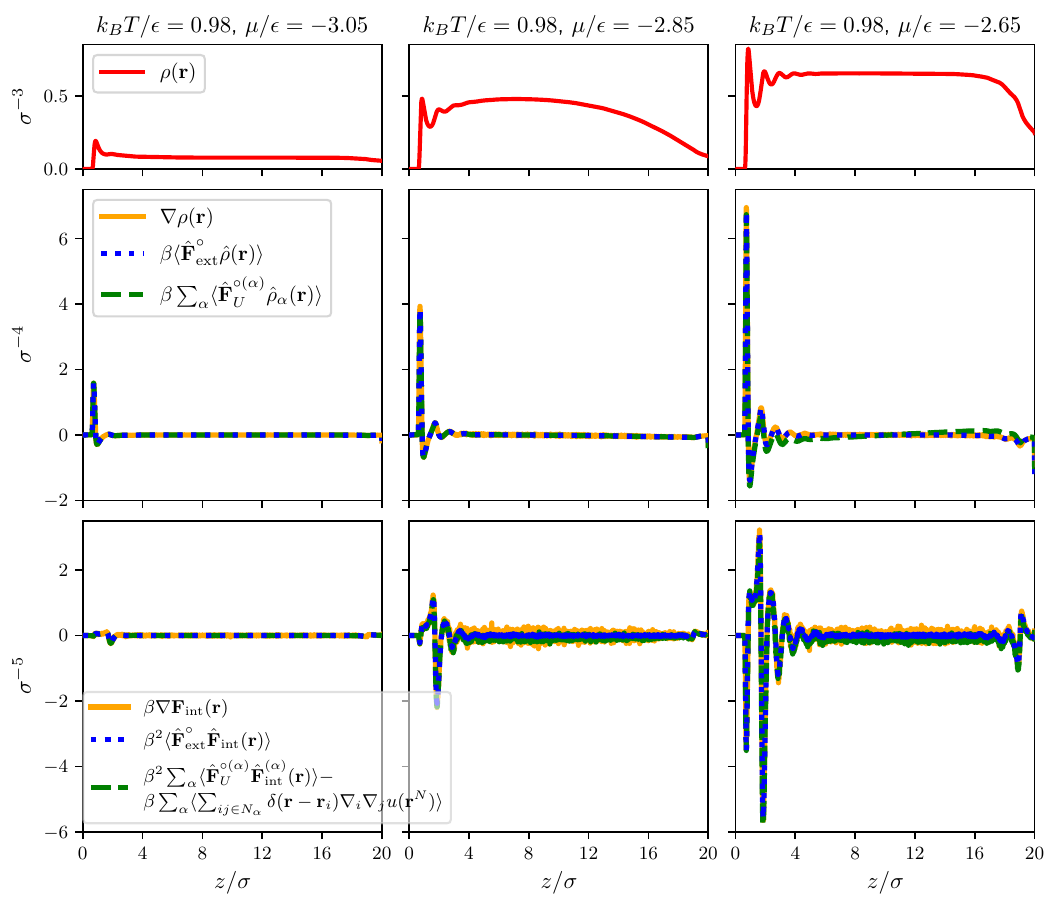}\\
  \hspace{7mm}
  \includegraphics[page=1,width=0.24\textwidth]{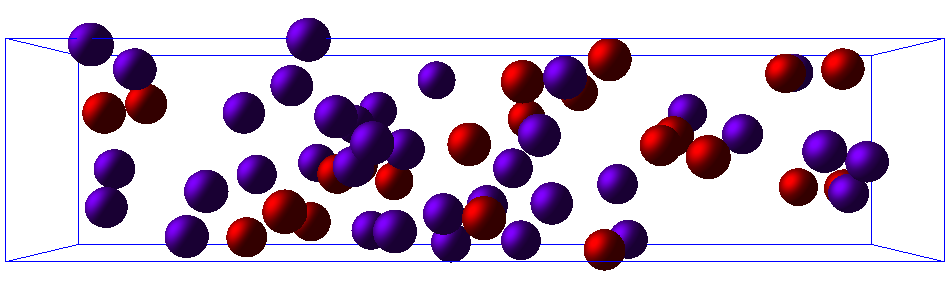}
  \includegraphics[page=1,width=0.24\textwidth]{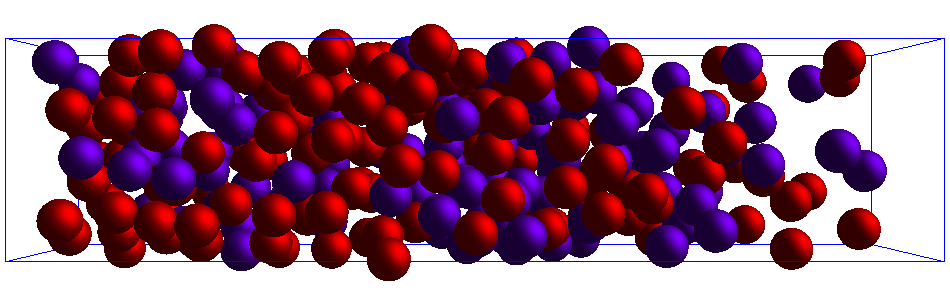}
  \includegraphics[page=1,width=0.24\textwidth]{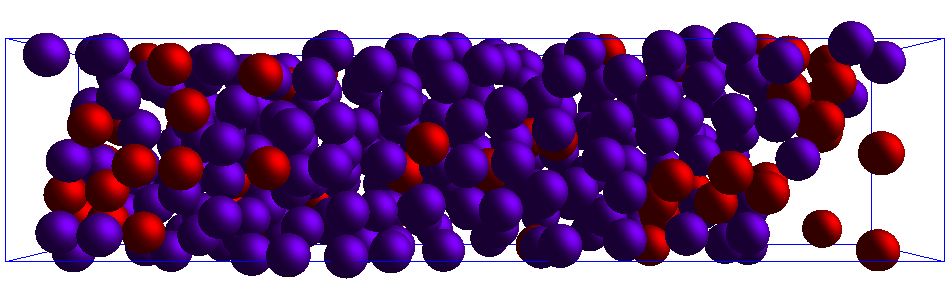}
  \caption{Specific hyperforce correlation functions of the
    symmetrical Lennard-Jones mixture of Wilding {\it et
      al.}~\cite{wilding1997, schmid2001wetting, wilding2002,
      wilding2003, wilding1998, koefinger2006epl, koefinger2006jcp} at
    scaled temperature $k_BT/\epsilon=0.98$ in the gas phase at
    chemical potential $\mu/\epsilon=-3.05$ (left column), in the
    mixed liquid phase at $\mu/\epsilon=-2.85$ (middle column), and in
    the demixed liquid phase at $\mu/\epsilon=-2.65$ (right
    column). The system is asymmetrically confined between an
    attractive wall (left) and a purely repulsive wall (right) and it
    is translationally invariant in the $x$ and $y$ directions.  Top
    row: the total density profile $\rho(\rv)=\rho_1(\rv)+\rho_2(\rv)$
    is shown as a function of the scaled distance $z/\sigma$ across
    the pore.
    Middle row: the gradient of the density profile, $\nabla\rho(\rv)
    = \partial \rho(z)/\partial z$, coincides numerically with the
    $z$-component of $\langle \hat
    \beta\hat\Fv_\rmext^\glob\hat\rho(\rv)\rangle$, according to
    \eqr{EQmixGlobalDensity2}, and with the $z$-component of
    $\sum_\alpha \langle \beta
    \Fv_U^{\glob(\alpha)}\hat\rho_\alpha\rangle$, according to
    \eqr{EQmixGlobalDensity3}.
    Bottom row: the $zz$-component of the gradient of the agglomerated
    local interparticle force density, $\nabla \beta \Fv_\rmint(\rv)$,
    coincides numerically with the $zz$-component of the correlation
    function $\langle \beta
    \hat\Fv_\rmext^\glob\hat\Fv_\rmint(\rv)\rangle$, according to
    \eqr{EQglobalSumRuleInterparticleForce1}, and with the
    $zz$-component of $\sum_\alpha\langle \beta \hat
    \Fv^{\glob(\alpha)}_{U} \beta\hat \Fv_{\rmint}^{(\alpha)}(\rv)
    \rangle - \sum_\alpha\langle \sum_{ij\in{\cal N}_\alpha}
    \delta(\rv-\rv_i)\nabla_i\nabla_j \beta u(\rv^N)\rangle$,
    according to \eqr{EQglobalSumRuleInterparticleForce2}.
    The bottom row displays corresponding simulation snashopts of
    the gas (left), mixed liquid (middle), and demixed liquid (right)
    states.}
  \label{FIGsilas1}
\end{figure*}

\section{Hyperforce sum rules for mixtures}
\label{SEChyperForceCorrelations}

\subsection{Local hyperforce sum rules}
\label{SEChyperForceGeneral}

We consider general observables $\hat A(\rv^N,\pv^N)$ and their
corresponding equilibrium average $A=\langle \hat A(\rv^N,
\pv^N)\rangle$. Following the argumentation of
Ref.~\cite{robitschko2024any}, the value of~$A$ is invariant under the
species-resolved shifting transformation
\eqref{EQmixtureCanonicalTransformationCoordinates} and
\eqref{EQmixtureCanonicalTransformationMomenta}. Hence the functional
derivative of the thermal average with respect to each shifting
field vanishes,
\begin{align}
  \frac{\delta A[\{\eps_\alphas\}]}{\delta \eps_\alpha(\rv)} &= 0.
  \label{EQAisInvariantMixture}
\end{align}
Using the explicit form of the equilibrium average as a phase space
integral, one can re-write \eqr{EQAisInvariantMixture} in the
following more explicit form:
\begin{align}
  -\beta\Big\langle\frac{\delta H[\{\eps_{\alpha'}\}]}{\delta \eps_\alpha(\rv)}
    \Big|_{\{\eps_\alphas=0\}} \hat A
    \Big\rangle
  +\Big\langle \frac{\delta\hat A[\{\eps_\alphas\}]}{\delta\eps_\alpha(\rv)}
  \Big|_{\{\eps_\alphas=0\}}\Big\rangle
  &= 0,
  \label{EQhyperforceSumRuleMixtureGeneric}
\end{align}
where all partial shifting fields $\{\eps_\alphas(\rv)\}$ have been
set to zero after the derivative is taken. Making the first term in
\eqr{EQhyperforceSumRuleMixtureGeneric} more explicit via its
relationship to the partial force density operator
\eqref{EQdeltaHdeltaEpsMixture} and also calculating the second term
explicitly leads to the following species-resolved hyperforce sum
rule:
\begin{align}
  \big\langle\beta\hatFva(\rv) \hat A \big\rangle 
  + \Big\langle
    \sum_{i\in {\cal N}_\alpha}\delta(\rv-\rv_i)\nabla_i\hat A
  \Big\rangle\;\qquad&
  \notag\\ 
  +\nabla\cdot\Big\langle
    \sum_{i\in {\cal N}_\alpha}
    \delta(\rv-\rv_i)\frac{\partial \hat A}{\partial \pv_i}\pv_i
    \Big\rangle &= 0,
  \label{EQhyperforceSumRuleMixture}
\end{align}
where the observable $\hat A(\rv^N,\pv^N)$ can have general phase
space dependence on $\rv^N,\pv^N$. Using the definition
\eqref{EQSvAdefinition} of the species-resolved hyperforce observable
$\hat \Sv_{A}^{(\alpha)}(\rv)$, we can put
\eqr{EQhyperforceSumRuleMixture} into the more compact form
\begin{align}
  \Sv_{A}^{(\alpha)}(\rv) +
  \langle \beta \hat \Fv_\alpha(\rv) \hat A \rangle
  &= 0,
  \label{EQhyperforceSumRuleMixtureCompactNotation}
\end{align}
where the partial mean hyperforce density is $\Sv_{A}^{(\alpha)}(\rv)=
\langle \hat \Sv_A^{(\alpha)}(\rv) \rangle$.

For cases where $\hat A$ is independent of the degrees of freedom of
species $\alpha$, as denoted by $\hat A(\{\rv_i, \pv_i\}_{i\notin{\cal
    N}_\alpha})$ the right hand side of the species-resolved
hyperforce density balance \eqref{EQhyperforceSumRuleMixture}
vanishes, which leads to the remarkably simple result:
\begin{align}
  \big\langle\beta\hatFva(\rv)
  \hat A(\{\rv_i,\pv_i\}_{i\notin{\cal N}_\alpha})\big\rangle &= 0.
  \label{EQhyperforceCrossSpecies}
\end{align}
Hence the total force density that acts on species $\alpha$ is
uncorrelated with all observables that only depend on the degrees of freedom of the species that
are different from $\alpha$. As the average partial force density
vanishes in equilibrium, $\Fva(\rv)=0$ according to
\eqr{EQforceDensityBalanceMixtureCompact}, the relationship
\eqref{EQhyperforceCrossSpecies} implies trivially that also
$\cov(\hatFva(\rv),\hat A(\{\rv_i,\pv_i\}_{i\notin{\cal
    N}_\alpha}))=0$.

In case that the considered observable is independent of momenta and
hence is a function of only the positions, i.e., $\hat A(\rv^N)$, the
last, momentum-dependent term on the left hand side of
\eqr{EQhyperforceSumRuleMixture} vanishes, and we obtain
\begin{align}
  \big\langle\beta \hatFva(\rv)\hat A(\rv^N) \big\rangle 
  +\Big\langle\sum_{i\in{\cal N}_\alpha}
    \delta(\rv-\rv_i)\nabla_i \hat A(\rv^N)
  \Big\rangle &= 0.
  \label{EQhyperforcePositionsMixture}
\end{align}
Using the standard splitting \eqref{EQforceDensityOperatorMixture} of
the position-dependent total force operator into interparticle,
external, and kinetic contributions allows one to re-write
\eqr{EQhyperforcePositionsMixture} in the following more explicit
form:
\begin{align}
  \big\langle\beta \hat \Fv_{\rmint}^{(\alpha)}(\rv)\hat A(\rv^N) \big\rangle
  -\big\langle\hat \rho_\alpha(\rv) \hat A(\rv^N) \big\rangle
  \nabla \beta V_{\rmext}^{(\alpha)}(\rv)\quad&
  \notag\\
  -\nabla \big\langle\hat\rho_\alpha(\rv)\hat A(\rv^N) \big\rangle
  + \avg{\sum_{i\in{\cal N}_\alpha}
    \delta(\rv-\rv_i)\nabla_i \hat A(\rv^N)} &= 0.
  \label{EQhyperforcePositionsMixture2}
\end{align}

For cases where the observable of interest depends only on the
position variables of an individual species $\alpha'$, as denoted by
$\hat A(\{\rv_i\}_{i\in {\cal N}_\alphas})$, we can particularize
\eqr{EQhyperforcePositionsMixture} further:
\begin{align}
 & \big\langle\beta \hatFva(\rv) \hat A(\{\rv_i\}_{i\in{\cal N}_\alphas})
  \big\rangle \notag\\&\qquad
  +\delta_\aas
  \Big\langle \sum_{i\in{\cal N}_\alpha}\delta(\rv-\rv_i)
  \nabla_i \hat A(\{\rv_i\}_{i\in {\cal N}_\alphas}) \Big\rangle = 0,
  \label{EQhyperforcePositionsMixture3}
\end{align}
such that the second term is only nonvanishing in the intraspecies
case, $\alpha=\alphas$.

\subsection{Global hyperforce sum rules}
\label{SEChyperForceGlobal}
Integrating \eqr{EQhyperforcePositionsMixture2} over position $\rv$
and exploiting that the diffusive gradient terms vanish in
systems enclosed by walls allows one to obtain the following global
hyperforce sum rule:
\begin{align}
  \big\langle\beta\hat\Fv_{\rmext}^{\glob(\alpha)}\hat A\big\rangle
  +\big\langle\beta\hat\Fv_{\rmint}^{\glob(\alpha)}\hat A \big\rangle
  + \avg{\sum_{i\in{\cal N}_\alpha}\nabla_i \hat A}
  &= 0.
  \label{EQglobalSumRule}
\end{align}
Here the global external force operator for species $\alpha$ is
defined as the sum over all external forces that act on this species,
i.e., $\hat\Fv_{\rmext}^{\glob(\alpha)}=-\sum_{i\in{\cal
    N}_\alpha}\nabla_i V_{\rmext}^{(\alpha)}(\rv_i)$ and
correspondingly for the species-resolved global interparticle force
$\hat \Fv_{\rmint}^{\glob(\alpha)} = -\sum_{i \in {\cal
    N}_\alpha}\nabla_i u(\rv^N)$.  Building the sum of both
contributions allows one to relate to the species-resolved global
potential force operator \eqref{EQglobalPotentialForceOperator} as
$\hat\Fv_U^{\glob(\alpha)} = \hat\Fv_\rmint^{\glob(\alpha)} +
\hat\Fv_\rmext^{\glob(\alpha)}$.  This allows one to rewrite
\eqr{EQglobalSumRule} in a more compact form,
\begin{align}
  \big\langle\beta\hat\Fv_U^{\glob(\alpha)}\hat A \big\rangle
  +\avg{\sum_{i\in{\cal N}_\alpha}\nabla_i \hat A}
  &= 0.
  \label{EQmixGlobalFU}
\end{align}
Note that the partial interparticle forces need not vanish
individually, yet when summed over all species $\sum_\alpha \hat
\Fv_{\rmint}^{\glob(\alpha)}=0$, as follows from Newton's third law
or, analogously, from the global translational invariance of the
interparticle potential $u(\rv^N)$ \cite{hermann2021noether}.

Summing the identity \eqref{EQglobalSumRule} over all species yields
\begin{align}
  \big\langle\beta\hat\Fv_{\rmext}^{\glob}\hat A\big\rangle
  + \avg{\sum_{i}\nabla_i \hat A}
  &= 0,
  \label{EQglobalSumRuleTotal}
\end{align}
where we have defined the global external force as
$\hat\Fv_{\rmext}^{\glob} =
\sum_\alpha\hat\Fv_{\rmext}^{\glob(\alpha)}$ and have exploited that
the global interparticle force operator vanishes, $\sum_\alpha
\hat\Fv_\rmint^{\glob(\alpha)}=0$, as described above.  We have
simplified the summation over particles in \eqr{EQglobalSumRuleTotal}
according to $\sum_i= \sum_\alpha\sum_{i\in{\cal N}_\alpha}$, where
the sum on the left hand side runs over all particles in the
system. The form of the sum rule \eqref{EQglobalSumRuleTotal} is
identical to the corresponding result for one-component systems
\cite{robitschko2024any}.

\subsection{Local sum rules for specific observables}
\label{SEChyperForceSpecific}

We apply the general sum rules described in
Sec.~\ref{SEChyperForceGeneral} above to several exemplary choices for
the `hyperobservable' $\hat A$, following the outline of the
single-component treatment in Ref.~\cite{robitschko2024any}.  As an
initial consistency check, the trivial choice $\hat A=1$ leads via
\eqr{EQhyperforcePositionsMixture3} [or alternatively via the simple
  \eqr{EQhyperforceCrossSpecies}] directly to the partial force
balance relationship \eqref{EQforceDensityBalanceMixture}.  Choosing
the partial density operator $\hat A =
\hat\rho_\alphas(\rv')=\sum_{i\in {\cal N}_\alphas}\delta(\rv'-\rv_i)$
and applying \eqr{EQhyperforcePositionsMixture3} yields
\begin{align}
  \big\langle\beta \hatFva(\rv)\hat\rho_\alphas(\rv') \big\rangle
  &= \delta_\aas
  \nabla' \rho_{2, \rm self}^{(\alpha)}(\rv,\rv'),
\end{align}
which is the density-force correlation sum rule
\eqref{EQsumRuleDensityForceMixture}. Taking the interparticle
potential energy $\hat A=u(\rv^N)$ leads via
\eqr{EQhyperforcePositionsMixture} to
\begin{align}
  \big\langle \beta \hatFva(\rv) u(\rv^N) \big\rangle
    &= \Fv_{\rm int}^{(\alpha)}(\rv).
\end{align}

Furthermore the (scaled) center of mass of all particles of species
$\alpha$, given by $\hat A=\sum_{i\in{\cal N}_\alphas}\rv_i$, leads
via \eqr{EQhyperforcePositionsMixture3} to the following identity:
\begin{align}
  -\avg{\beta \hatFva(\rv)\sum_{i\in{\cal N}_\alphas} \rv_i}
  &= \delta_\aas\unity \rho_\alpha(\rv),
\end{align}
where we recall $\unity$ as the $d\times d$-unit matrix with $d$
indicating the spatial dimensionality.

\subsection{Global sum rules for specific observables}
\label{EQspecifcGlobalSumRules}

We formulate several global sum rules that arise from making specific
choices of hyperobservables. Our aim below in
Sec.~\ref{SECsimulationResults} will be to demonstrate in simulations
the validity of these sum rules and to show the accessibility of the
correlation functions that are involved.

We first address the partial density operator.  In the general sum
rule \eqref{EQmixGlobalFU} we set $\hat A = \hat\rho_{\alpha'}(\rv)$
and use the simplification $\langle\sum_{i\in{\cal N}_\alpha}\nabla_i
\hat \rho_\alphas(\rv) \rangle = -\delta_\aas \nabla
\rho_\alpha(\rv)$, where we recall that no summation over double
species indices is implied in our notation. Then one obtains the
following sum rule:
\begin{align}
  \big\langle\beta\hat\Fv_U^{\glob(\alpha)}\hat \rho_\alphas(\rv)\big\rangle
  -\delta_\aas \nabla \rho_\alpha(\rv)
  &= 0,
  \label{EQmixGlobalDensity1}
\end{align}
which applies to all combinations of $\alpha,\alphas$. Summing
\eqr{EQmixGlobalDensity1} over all pairs of species yields
\begin{align}
  \big\langle \beta \hat\Fv^{\glob}_{\rmext}\hat\rho(\rv)\big\rangle
  -\nabla\rho(\rv) &= 0,
  \label{EQmixGlobalDensity2}
\end{align}
which is analogous in form to the one-component hyperforce sum rule
for the density operator \cite{robitschko2024any}.  The agglomerated
density operator is obtained by summing over all species according to
$\hat\rho(\rv) = \sum_\alpha \hat\rho_\alpha(\rv)$ and we recall
$\hat\Fv^{\glob}_\rmext$ as the agglomerated external force.

Specializing \eqr{EQmixGlobalDensity1} to the case $\alpha=\alphas$
and summing over the remaining species index gives the following
alternative form:
\begin{align}
  \Big\langle \sum_\alpha
  \beta\hat\Fv_{U}^{\glob(\alpha)}\hat\rho_\alpha(\rv)\Big\rangle -
  \nabla\rho(\rv) &= 0,
  \label{EQmixGlobalDensity3}
\end{align}
where the first (correlation) term differs from that in
\eqr{EQmixGlobalDensity2}: i) interparticle interactions contribute
and ii) only intraspecies correlations between forces and densities
occur. Equations \eqref{EQmixGlobalDensity2} and
\eqref{EQmixGlobalDensity3} are both suitable for situations where one
is interested in the behaviour of the agglomerated density profile
$\rho(\rv)$ rather than its partial variants $\rho_\alpha(\rv)$, as
can be advantageous in situations of demixing phase separation.

We next consider the species-resolved interparticle force density
operator $\hat A = \beta \hat\Fv_\rmint^{(\alphas)}(\rv)$, of which we
recall the explicit form $\beta\hat\Fv_\rmint^{(\alphas)}(\rv) =
-\sum_{i\in{\cal N}_{\alphas}}\delta(\rv-\rv_i)\nabla_i \beta
u(\rv^N)$. Then specializing the global sum rule \eqref{EQmixGlobalFU}
yields
\begin{align}
  \big\langle\beta\hat\Fv_U^{\glob(\alpha)}
  \beta\hat\Fv_\rmint^{(\alphas)}(\rv)\big\rangle
  +\avg{\sum_{i\in{\cal N}_\alpha}\nabla_i 
    \beta\hat\Fv_\rmint^{(\alphas)}(\rv)}
  &= 0.
  \label{EQmixGlobalForce1}
\end{align}
Note that $\hat \Fv_\rmint^{(\alpha')}(\rv)$ will in general depend
also on further species $\alpha'\neq\alpha$, such that the gradient in
the second term in \eqr{EQmixGlobalForce1} will in general be nonzero;
this situation is different from the mechanism in
\eqr{EQhyperforcePositionsMixture3} where $\hat A(\{\rv_i\}_{i\in{\cal
    N}_\alphas})$ depends {\it solely} on the positions of species
$\alpha'$.

Furthermore we address the thermally scaled agglomerated interparticle
force density $\hat A = \beta\hat\Fv_\rmint(\rv) = \sum_\alpha\hat
\beta\Fv_\rmint^{(\alpha)}(\rv)$. Then from \eqr{EQglobalSumRuleTotal}
and Newton's third law one obtains the following global sum rule:
\begin{align}
   \big\langle \beta \hat \Fv^{\glob}_{\rmext}
  \beta \hat \Fv_{\rmint}(\rv)\big\rangle
  - \nabla \beta \Fv_{\rmint}(\rv)
  &= 0.
  \label{EQglobalSumRuleInterparticleForce1}
\end{align}

One obtains an alternative to \eqr{EQglobalSumRuleInterparticleForce1}
by first specializing \eqr{EQmixGlobalForce1} to the case of equal
species, $\alpha=\alphas$, and then building the sum over the
remaining joint species index. Simplifying the result yields:
\begin{align}
  & \sum_\alpha\big\langle \beta \hat \Fv^{\glob(\alpha)}_{U}
  \beta\hat \Fv_{\rmint}^{(\alpha)}(\rv)
  \big\rangle \notag\\
  & - \sum_\alpha\Big\langle
  \sum_{ij\in{\cal N}_\alpha} \delta(\rv-\rv_i)\nabla_i\nabla_j \beta u(\rv^N)
  \Big\rangle
  -\nabla \beta\Fv_{\rmint}(\rv)
  = 0.
  \label{EQglobalSumRuleInterparticleForce2}
\end{align}
Hence both sum rules \eqref{EQglobalSumRuleInterparticleForce1} and
\eqref{EQglobalSumRuleInterparticleForce2} feature the
thermally-scaled negative gradient of the agglomerated interparticle
force density $-\nabla\beta \Fv_\rmint(\rv)$ as the last term on the
left hand sides.

We test the sum rules \eqref{EQmixGlobalDensity2},
\eqref{EQmixGlobalDensity3},
\eqref{EQglobalSumRuleInterparticleForce1}, and
\eqref{EQglobalSumRuleInterparticleForce2} in simulations, as
described in the following.

\section{Simulation results}
\label{SECsimulationResults}

To illustrate the gauge correlation theory we apply it to a concrete
system and hence consider the prototypical binary Lennard-Jones
mixture in three spatial dimensions. The system is characterized by
pair potentials $\phi_\aas(r)$ that act between particles of species
$\alpha$ and $\alphas$. The species-labelled Lennard-Jones potential
is thereby given by
\begin{align}
  \phi_\aas(r) &=
  4\epsilon_\aas 
  \Big[
    \Big(\frac{\sigma_\aas}{r}\Big)^{12}
    -\Big(\frac{\sigma_\aas}{r}\Big)^{6}
    \Big],
  \label{EQphiaasGeneral}
\end{align}
where $r$ is the separation distance between the two particles,
$\epsilon_\aas$ are energy parameters, and $\sigma_\aas$ are
lengthscales, with the species indices taking on values
$\alpha,\alphas=1,2$ in a two-component system and implying the
symmetry $\phi_{12}(r) = \phi_{21}(r)$.

\subsection{Bulk force-force correlation structure}
\label{SECbulkStructure}

We first address the pair gauge correlation structure of a bulk liquid
using the Lennard-Jones parameterization due to Kob and Andersen
\cite{kob1994, kob1995}.  The fundamental length scale $\sigma$ and
energy scale $\epsilon$ are taken to be those of the first component
$\alpha=1$, such that $\sigma_{11} = \sigma$ and $\epsilon_{11} =
\epsilon$. The intraspecies interactions amongst particles of
species~$\alpha=2$ are characterized by a smaller length scale,
$\sigma_{22} = 0.88\sigma$, and weakened energy scale, $\epsilon_{22}
= 0.5\epsilon$. The cross species length scale is reduced,
$\sigma_{12} = 0.8 \sigma$, and the cross interactions are
strengthened, $\epsilon_{12} = 1.5\epsilon$, as compared to both
intraspecies pair potentials.

We use adaptive Brownian dynamics \cite{sammueller2021} to sample the
gauge correlation functions of the bulk liquid at thermal equilibrium
in the canonical ensemble.  We use a cubic simulation box with lateral
size $10\sigma$ and total particle number $N=844$. Hence the total
bulk density is $\rho_b = N/V= 0.844\sigma^{-3}$ and the partial bulk
densities are $\rho_1^b\sigma^3=0.591$ and $\rho_2^b\sigma^3=0.253$,
and we choose the temperature as $k_BT/\epsilon=1.1$. We truncate all
pair potentials at a cutoff distance $r_c=2.5\sigma$ and potential
shifts are applied such that each $\phi_\aas(r)$ is continuous at
$r_c$.  We have used an initial simulation period of temporal length
$5 \tau_B$ for equilibration, with Brownian time scale
$\tau_B=\sigma^2 k_BT/(D_0 \epsilon)$, where $D_0$ is the
single-particle diffusion constant. The data is collected over 25
runs, which amounts to an overall time $4000\tau_B$ that consists of
$\sim 6.5\cdot 10^7$ adaptive time steps \cite{sammueller2021}. The
parameters for the adaptive Brownian dynamics tolerance criterium are
set as 0.1 (relative tolerance) and 0.01 (absolute tolerance); see
Ref.~\cite{sammueller2021} for details.

In Fig.~\ref{FIGgofr} we display results for the partial pair
distribution functions, for the force-gradient correlation functions,
and for the force-force correlation functions.  The partial pair
distribution functions $g_{\aas}(r)$ display pronounced spatial
structuring that is typical of the liquid state, see the first row in
Fig.~\ref{FIGgofr}. Both $g_{11}(r)$ and $g_{12}(r)$ possess
pronounced first peaks, which are indicative of the formation of
nearest neighbour coordination shells. The subsequent decay for
increasing distance $r/\sigma$ is damped oscillatory, on the linear
scale considered here. The first peak of $g_{22}(r)$ is less strongly
pronounced than for the 11- and 12-pairs, but the decay towards larger
values of $r/\sigma$ is also damped oscillatory, as is expected from
the general theory of asymptotic decay of correlations in liquids
\cite{evans1993decay, evans1994decay, dijkstra2000decay,
  grodon2004decay, cats2021decay}, which ascertains common type of
decay for all partial pair distribution functions.  The agglomerated
pair distribution function $g(r)$ displays rich oscillatory structure.
This structuring arises from the linear combination of the underlying
partial contributions, see the fourth panel in the first row of
Fig.~\ref{FIGgofr}.

Results for the force-gradient correlation function ${\sf
  g}_\gradf^{(\aas)}(r)$ are shown in the second and third row of
Fig.~\ref{FIGgofr}, where we display, respectively, the parallel and
the perpendicular tensor components. Simulation results are obtained
by sampling the force gradients via finite differences, which are
built by performing virtual particle displacements
\cite{sammueller2023whatIsLiquid}.  This `direct' method is of
universal applicability to general many-body interparticle
interactions, such as the monatomic water model \cite{molinero2009}
and the three-body gel former \cite{saw2009, saw2011,
  sammueller2023gel}, both of which are special parameter choices of
the general Stillinger-Weber model~\cite{stillinger1985}.  The finite
difference method also circumvents the need to implement Hessians of
interparticle potentials explicitly.

For the present pairwise-interacting mixture, each partial
force-gradient pair correlation function is related to the
corresponding partial pair distribution function $g_\aas(r)$ and the
Hessian of the pair potential $\nabla \nabla\phi_\aas(r)$, see
\eqr{EQforceGradientViaPairPotential} and the dotted lines in the
second and third row of Fig.~\ref{FIGgofr}. The identity
\eqref{EQforceGradientViaPairPotential} is verified numerically, see
Fig.~\ref{FIGgofr}.  The parallel and perpendicular tensor components
display respectively a strong positive and negative first peak, as can
be expected from the underlying second and first derivatives of the
respective pair potential with respect to distance. Both tensor
components strictly vanish beyond the potential truncation distance
$r_c$. The perpendicular component displays a positive overshoot for
each pair of species.  We attribute the effect to the presence of
interparticle attraction, see Ref.~\cite{sammueller2023gel} for a
comparison of results for the one-component Lennard-Jones fluid and
for the purely repulsive (also one-component) Weeks-Chandler-Andersen
fluid, where in the latter model the effect is absent.

Results for the partial force-force pair correlation functions ${\sf
  g}_\ff^{(\aas)}(r)$ are shown in the fourth and fifth row of
Fig.~\ref{FIGgofr}.  We recall that this correlation function measures
the correlation of the sum of all interparticle forces acting on each
particle of the considered pair. No simplification arises that would
be similar to that for the above force gradient correlation function
${\sf g}_\gradf^{(\aas)}(r)$. Performing the sampling of the
force-force gradient correlation function is straightforward, in
particular when using methods that already provide direct access to
forces, as is the case for the present adaptive Brownian dynamics or,
similarly, in Molecular Dynamics. A description of a suitable choice
of coordinate system that facilitates straightforward access to the
parallel and perpendicular tensor components is given in
Ref.~\cite{sammueller2023whatIsLiquid}.

The partial force-force correlation functions again display rich
spatial structuring. The parallel component has a strong first
negative peak, as is indicative of anti-correlated forces on two
mutually interacting particles. Note that the negative sign is
indicative of anticorrelation both for repulsion in the core region
and for the longer-ranged attraction. The perpendicular tensor
component has smaller amplitude and smoother variation with distance.
The two tensor components, see
Eqs.~\eqref{EQthreegSumRulePairParallel} and
\eqref{EQthreegSumRulePairPerp}, of the species-resolved 3g-sum rule
\eqref{EQgIdentityMix}, as well as the species-agglomerated sum rules,
see Eqs.~\eqref{EQgIdentityAgglomeratedParallel} and
\eqref{EQgIdentityAgglomeratedPerpendicular}, are satisfied to
excellent accuracy.  We hence conclude that the gauge correlation
framework offers significant and physically meaningful insight into
the spatial structure of bulk liquid mixtures.

\subsection{Confinement between parallel walls}
\label{SECconfinement}

To consider a second model fluid, we follow Wilding {\it
  et~al.}~\cite{wilding1997, schmid2001wetting, wilding1998,
  wilding2002, wilding2003, koefinger2006epl, koefinger2006jcp}, who
investigated a symmetrical Lennard-Jones mixture with a single common
lengthscale $\sigma_{11}=\sigma_{22}=\sigma_{12}=\sigma$. The two
intraspecies interaction strengths are identical,
$\epsilon_{11}=\epsilon_{22}=\epsilon$, and the cross-species
interaction strength is weakened by comparison,
$\epsilon_{12}=0.7\epsilon$. The cutoff radius is again chosen as $r_c
= 2.5 \sigma$ and no potential shift is applied.  We use grand
canonical Monte Carlo simulations \cite{frenkel2023book, wilding2001,
  brukhno2021dlmonte} to generate equilibrium data.  The system
exhibits intricate phase behaviour, including gas-liquid and
liquid-liquid phase coexistence phenomena, which were recently
re-addressed using neural density functional learning
\cite{robitschko2025mixShort}.

We demonstrate the applicability of the hyperforce sum rules to spatially
inhomogeneous systems by considering confinement in a planar asymmetric slit
pore. The left wall is thereby taken to be of Lennard-Jones 9--3 type
\cite{schmid2001wetting}: $V_{\rmext,L}(z)=\epsilon [(2/15)(\sigma/z)^9 -
(\sigma/z)^3]$, where $z$ measures the distance from the wall and we recall
$\epsilon$ as the common energy scale. The right wall is purely repulsive:
$V_{\rmext,R}(z) = 4 \epsilon_w [(\sigma/z)^{12}-(\sigma/z)^6+1/4]$, where the
wall potential strength is taken to be $\epsilon_w=10\epsilon$, and the cutoff
is at $z_c= 2^{1/6} \sigma$. The total external potential is then given by the
combination $V_\rmext(z)=V_{\rmext,L}(z) + V_{\rmext,R}(L_z-z)$, with simulation
box size $L_z$ in the $z$-direction. We choose $L_z=20\sigma + 2^{1/6}\sigma =
21.122 \sigma$ and take the box size in the two lateral directions to
be~$5\sigma$. For each of the three statepoints considered we have carried out
$2\cdot 10^9$ (gas) and $3\cdot 10^9$ (mixed and demixed liquid) grand canonical
Monte Carlo single particle moves. Each move consists of either a position
displacement with maximal length $0.2\sigma$ or particle insertion/deletion
attempt, performed with probability 0.1. We collect data after an equilibration
period of $3\cdot 10^5$ Monte Carlo steps.

Fig.~\ref{FIGsilas1} shows our simulation results for the specific
hyperforce correlation functions laid out in
Sec.~\ref{EQspecifcGlobalSumRules}. These correlation functions are
chosen specifically to facilitate access both to the gradient of the
density profile and to the gradient of the localized interparticle
force density. We hence consider the sum rules
\eqref{EQmixGlobalDensity2} and \eqref{EQmixGlobalDensity3} for the
gradient of the total density profile, and the identities
\eqref{EQglobalSumRuleInterparticleForce1} and
\eqref{EQglobalSumRuleInterparticleForce2} for the gradient of the
agglomerated interparticle force density. We choose three statepoints,
as typical for the gas, the mixed liquid, and the demixed liquid
phase. The comparisons shown in Fig.~\ref{FIGsilas1} demonstrate that
in all cases considered, we find the sum rules to be satisfied.

The total density profile, $\rho(z)=\rho_1(z) + \rho_2(z)$ as shown in
the first row of Fig.~\ref{FIGsilas1}, is indicative of capillary
structuring in the low-density phase (left column), in the mixed
liquid (middle column) and in the demixed liquid (right column). The
respective simulation snapshots illustrate these different capillary
states. The gradient of the total density profile, $\nabla\rho(z)$,
shows pronounced oscillations at the (left) attractive wall, where
molecular packing effects are apparent. This feature increases upon
increasing values of $\mu/\epsilon$ (from left to right in
Fig.~\ref{FIGsilas1}). The gradient of the interparticle force density
(third row in Fig.~\ref{FIGsilas1}) shows even more pronounced
structuring than the density gradient for the densest system
considered. The caption of Fig.~\ref{FIGsilas1} gives details about
the specific sum rules that are demonstrated.

\section{Conclusions}
\label{SECconclusions}

In summary, we have explored the consequences of gauge invariance with
respect to species-resolved phase space shifting in multi-component
classical many-body systems. The gauge transformation constitutes a
species-specific canonical transformation that acts on the fundamental
position and momentum degrees of freedom. The transformation is
represented by differential operators that act on general phase space
observables and that feature Lie algebra commutator structure. The
geometric nature of the gauge transformation, see
Ref.~\cite{mueller2024whygauge} for an in-depth description, renders
the framework generally applicable to arbitrary phase space functions
$\hat A$ as the `hyperobservable' of interest. The hyperobservable can
be a bespoke order parameter that is relevant for the physics of the
system under consideration or, alternatively, it can be chosen as a
more standard observable, such as the partial one-body density and
force density observables, as we have considered here in our
simulation work.

We have described the rich formal gauge correlation structure that
emerges in mixtures, where the hyperobservable, in analogy to the
Hamiltonian itself \cite{sammueller2024whyhyperDFT}, generates
corresponding spatially-resolved hyperforce density observables via
phase space differentiation. Upon building the thermal equilibrium
average, the mean partial one-body hyperforce density is related, in a
formally exact way, via equilibrium sum rules to the correlation of
the hyperobservable with the spatially localized force density. We
have described in detail several relevant special cases of these sum
rules that are relevant for mixtures, such as, e.g., arising from the
absence of the dependence on specific species, which leads to formal
simplification.

Turning to the two-body level of the gauge correlation functions, we
have generalized the emerging two-body force-force and force-gradient
correlation framework, as were formulated originally for pure systems
\cite{sammueller2023whatIsLiquid, hermann2023whatIsLiquid}, to
multi-component systems. Our corresponding simulation work for the
Kob-Andersen model, in its liquid phase, has shown that deep insights
into the liquid structure can be gained specifically from the partial
force-force and force-gradient correlation functions. We found all
corresponding sum rules to be satisfied numerically. We recall that
their validity hinges on thermal equilibrium. Hence our present
investigation could serve as a platform to shed new light on the rich
topic of the nonequilibrium nature of glasses that arise from
performing a temperature quench of the liquid. We refer to
Ref.~\cite{mueller2024dynamic} for the formulation of dynamical
hypercurrent sum rules that arise from the dynamical generalization of
phase space shifting.
In future work it would be interesting to explore possible
connections of the present framework with mode-coupling theory
\cite{janssen2018}. We also leave the investigation of the gauge
correlation sum rules for systems interacting with multi-body
interparticle interactions to future work.

We have demonstrated the applicability of the gauge correlation
framework to spatially inhomogeneous systems by carrying out
simulations for a symmetrical Lennard-Jones system previously
investigated by Wilding~{\it et~al.}~\cite{wilding1997,
  schmid2001wetting, wilding2002, wilding2003, wilding1998,
  koefinger2006epl, koefinger2006jcp} in the context of bulk phase
behaviour and associated interfacial physics. We have ascertained that
both the gradient of the density profile and the gradient of the
one-body force density distribution are accessible via corresponding
hyperforce correlation functions, see Fig.~\ref{FIGsilas1} and the
description given in Sec.~\ref{SECconfinement}.

Concerning phase-separating systems, it would be relevant to consider
specific hyperobservables that would relate to near-coexistence
conditions, such as the local compressibilities reflecting density
fluctuations. Note that local fluctuation profiles are closely
connected with a corresponding local compressibility, when choosing
the hyperobservable as $\hat A = N_\alpha$. However, due to the
significant flexibility in choosing $\hat A$, we can envisage much
potential for shedding new light on phase-separating systems as well
as on the nature of ordered phases, such as crystalline solids.

We have used two specific parameterizations of the binary
Lennard-Jones systems to exemplify our framework. For similar and
related parameter choices, a wealth of relevant research questions has
been addressed. This includes transport phenomena in mixtures
\cite{das2003}, critical dynamics and finite-size scaling
\cite{das2006prl, das2006jcp, roy2011}, phase separation inside of
nanopores \cite{basu2016}, the structure and dynamics near demixing
\cite{roy2016}, sub-system analysis \cite{pathania2021}, the study of
hydrodynamic effects \cite{das2023}, spinodal decomposition
\cite{zaidi2024}, as well as critical surface adsorption
\cite{roy2025}.  A further important class of models consist of
depletion-based binary mixtures, where an added secondary (depletion)
agent generates an effective interaction between the primary
(colloidal) component \cite{lekkerkerker2024book}.

The fact that the hyperobservable can be of very general nature allows
one to address concrete applications in flexible ways. For further
specific examples, we refer to Refs.~\cite{sammueller2024hyperDFT,
  sammueller2024whyhyperDFT} for investigations of the
hyperfluctuation profile that is associated with a clustering order
parameter and to Refs.~\cite{eckert2020, eckert2023fluctuation,
  coe2022pre} for the local thermal susceptibility, which arises from
addressing the entropy.  We re-iterate that choosing the
species-resolved number of particles, $\hat A = N_\alpha$, leads to
partial versions of the local compressibility \cite{stewart2012pre,
  evans2015jpcm, evans2019pnas, coe2022prl, coe2022pre}.  In future
work, it would be interesting to investigate connections to the
reduced-variance (force-sampling) \cite{borgis2013,
  delasheras2018forceSampling, rotenberg2020, coles2021} and mapped
averaging \cite{schultz2016, purohit2019, moustafa2022} schemes.
In particular the two-body framework of Sec.~\ref{SECthreegRules}
offers potential for such use; note that
Eqs.~\eqref{EQthreegSumRulePairParallel} and
\eqref{EQthreegSumRulePairPerp} can be viewed as differential
equations for the partial pair distribution functions, provided that
(simulation) results for the partial force-force correlation functions
are available.

The present theory can form an important role in neural functional
construction \cite{delasheras2023perspective, sammueller2023neural,
  sammueller2023whyNeural, sammueller2024hyperDFT,
  sammueller2024whyhyperDFT, sammueller2024pairmatching,
  zimmermann2024ml, sammueller2024attraction, buchannan2025attraction,
  kampa2024meta, sammueller2025chemicalPotential,
  robitschko2025mixShort}, as the sum rules carry significant
potential for serving as diagnostic tools to assess the
self-consistency of numerical predictions.
The hyperforce gauge correlation identities tie in particularly well
with the hyperdensity functional framework for the behaviour of
general observables in spatially inhomogeneous systems
\cite{sammueller2024hyperDFT, sammueller2024whyhyperDFT}, as follows
from the fundamental Mermin-Evans density functional map
\cite{mermin1965, evans1979}.

Sum rules are of significant importantance in first-principles-based
machine learning in soft matter physics
\cite{delasheras2023perspective, sammueller2023neural,
  sammueller2023whyNeural, sammueller2024hyperDFT,
  sammueller2024whyhyperDFT, sammueller2024pairmatching,
  zimmermann2024ml, sammueller2024attraction, buchannan2025attraction,
  kampa2024meta, sammueller2025chemicalPotential,
  robitschko2025mixShort}, as also applied to charged
\cite{bui2024neuralrpm, bui2025electromechanics,
  bui2025dielectrocapillarity} and further \cite{kelley2024ml,
  simon2023mlPatchy, simon2024patchy, dijkman2024ml} relevant model
systems. The basis for the neural functional learning method
\cite{sammueller2023neural, sammueller2023whyNeural} are the formally
exact functional relationships provided by classical density
functional~\cite{evans1979} and power functional theory
\cite{schmidt2022rmp}. Statistical mechanical sum rules can serve as
systematic means to assess the quality of the neural predictions and
as regularizers during training.
In particular, the method of local learning, which applies to either
the one-body direct correlation functional, to the excess free energy
functional \cite{sammueller2024pairmatching}, or the local
nonequilibrium force density \cite{delasheras2023perspective,
  zimmermann2024ml}, incorporates the effects of the interparticle
interactions in a highly efficient, functional form.

\vspace{5mm}
\begin{center}
  {\bf Data availability}
\end{center}
Simulation data is openly available at Zenodo
\cite{matthes2024mixZenodo}.

\bigskip

\begin{acknowledgments}
  We thank Bob Evans and Nigel Wilding for useful discussions. This
  work is supported by the DFG (Deutsche Forschungsgemeinschaft) under
  Project No.~551294732. Some of the calculations were performed using
  the emil-cluster of the Bayreuth Centre for High Performance
  Computing funded by the DFG (Deutsche Forschungsgemeinschaft) under
  Project No.~422127126.  S.~H.\ received funding from the European
  Union under the Horizon Europe Framework for Research and Innovation
  via the Marie Sk\l odowska-Curie Grant 101149232, Hyperion.
\end{acknowledgments}


\end{document}